\documentclass[final,5p,times,twocolumn]{elsarticle}
\usepackage{hyperref}
\usepackage{lineno,color}
\modulolinenumbers[3]
\journal{Astroparticle Physics}
\usepackage{amsmath}
\usepackage{graphicx}
\usepackage{epstopdf}
\usepackage{fancyhdr}
\usepackage[normalem]{ulem}
\usepackage{booktabs}
\newcommand\apj{ApJ}

\newcommand\prl{Phys. Rev. Lett.}
\newcommand\prd{Phys. Rev. D}

\def\procspie{\ref@jnl{Proc.~SPIE}}   

\usepackage{soul}
\usepackage{float}
\usepackage{fixltx2e}

\begin{document}                

\begin{frontmatter}
  \title{\vspace{-1.9cm}Limits  on the  flux of tau neutrinos from  1 PeV  to  3 EeV   with the MAGIC telescopes  }
\author[1]{\vspace{-0.2cm}M.~L.~Ahnen}
\author[2,20]{S.~Ansoldi}
\author[3]{L.~A.~Antonelli }
\author[4]{C.~Arcaro }
\author[5]{D.~Baack}
\author[6]{A.~Babi\'c }
\author[7]{B.~Banerjee}
\author[8]{P.~Bangale }
\author[8,9]{U.~Barres de Almeida }
\author[10]{J.~A.~Barrio}
% ...

\address[1]{ETH Zurich, CH-8093 Zurich, Switzerland}
\address[2]{ Universit\`a di Udine, and INFN Trieste, I-33100 Udine, Italy}
\address[3]{ National Institute for Astrophysics (INAF), I-00136 Rome, Italy}
\address[4]{ Universit\`a di Padova and INFN, I-35131 Padova, Italy}
\address[5]{Technische Universit\"at Dortmund, D-44221 Dortmund, Germany}
\address[6]{ Croatian MAGIC Consortium: University of Rijeka, 51000 Rijeka, University of Split - FESB, 21000 Split,
University of Zagreb - FER, 10000 Zagreb, University of Osijek, 31000 Osijek and Rudjer Boskovic Institute, 10000
Zagreb, Croatia.}
\address[7]{ Saha Institute of Nuclear Physics, HBNI, 1/AF Bidhannagar, Salt Lake, Sector-1, Kolkata 700064, India}
\address[8]{ Max-Planck-Institut f\"ur Physik, D-80805 M\"unchen, Germany}
\address[9]{ now at Centro Brasileiro de Pesquisas F\'isicas (CBPF), 22290-180 URCA, Rio de Janeiro (RJ), Brasil}
\address[10]{ Unidad de Part\'iculas y Cosmolog\'ia (UPARCOS), Universidad Complutense, E-28040 Madrid, Spain}
% ...
\address[20]{Japanese MAGIC Consortium: ICRR, The University of Tokyo, 277-8582 Chiba, Japan; Department of Physics, Kyoto University, 606-8502 Kyoto, Japan; Tokai University, 259-1292 Kanagawa, Japan; The University of Tokushima, 770-8502 Tokushima, Japan}
% ...
\cortext[mycorrespondingauthor]{Corresponding authors: Dariusz Gora (dariusz.gora@ifj.edu.pl) and Elisa Bernardini (Elisa.Bernardini@desy.de)}

\begin{abstract}
A search for tau neutrino induced showers with the MAGIC telescopes is presented. The MAGIC telescopes  located at an altitude of 2200 m a.s.l. in the Canary Island of La Palma,
can   point   towards the  horizon or a few degrees below across an  azimuthal range of about 80 degrees. This  provides  a possibility to search for  air showers induced by  tau leptons 
arising from interactions of tau neutrinos in the Earth crust or the  surrounding ocean. In this paper we show  how such air showers can be discriminated from the background of very inclined hadronic showers by using Monte Carlo simulations. Taking into account the orography of the site, the point source acceptance and the event rates expected have been calculated for a sample of generic neutrino fluxes from photo-hadronic interactions in AGNs. The analysis of about 30 hours of data taken towards the sea
leads to a  90\% C.L.  point source limit for tau neutrinos in the energy range from $1.0  \times 10^{15}$ eV 
to $3.0  \times 10^{18}$ eV of about $E_{\nu_{\tau}}^{2}\times \phi (E_{\nu_{\tau}}) < 2.0 \times 10^{-4}$ GeV cm$^{-2}$ s$^{-1}$ for an assumed power-law neutrino spectrum with spectral index $\gamma$=-2. However, with 300 hours 
and in case of an optimistic neutrino flare model, limits of the level down to
   $E_{\nu_{\tau}}^{2}\times \phi (E_{\nu_{\tau}}) < 8.4 \times 10^{-6}$ GeV cm$^{-2}$ s$^{-1}$ can be expected. 
\end{abstract}

\end{frontmatter}

\section{Introduction}
The discovery of an astrophysical flux of high-energy neutrinos by IceCube~\cite{HESE2} was a major step 
forward in the ongoing search for the origin of cosmic rays, since  neutrino emission 
needs to be  produced by hadronic interactions in astrophysical accelerators. 
The observed neutrino flux  by IceCube  and its compostion is in agreement with equal fractions of all neutrino flavours~\cite{icecuflavour,icecuflavour1}. Tau neutrinos in the IceCube  flux should also be  expected, due to  neutrino  oscillation, but up to now, $\nu_{\tau}$  have not been identified.

\begin{figure*}[h]
\vspace{-0.2cm}
\includegraphics [width=0.40\textwidth, height=8.5cm]{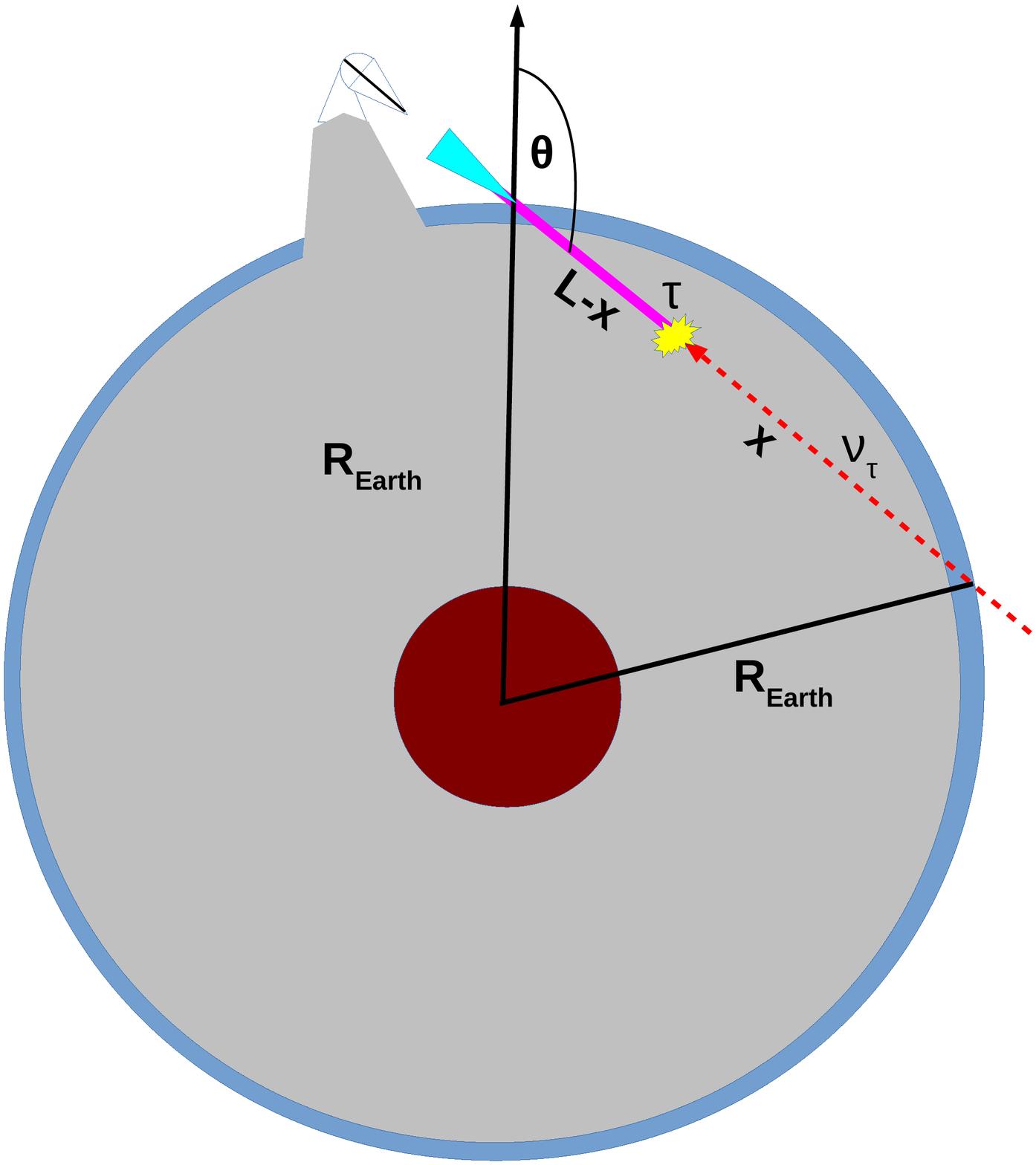}
\includegraphics [width=0.55\textwidth,height=6.5cm]{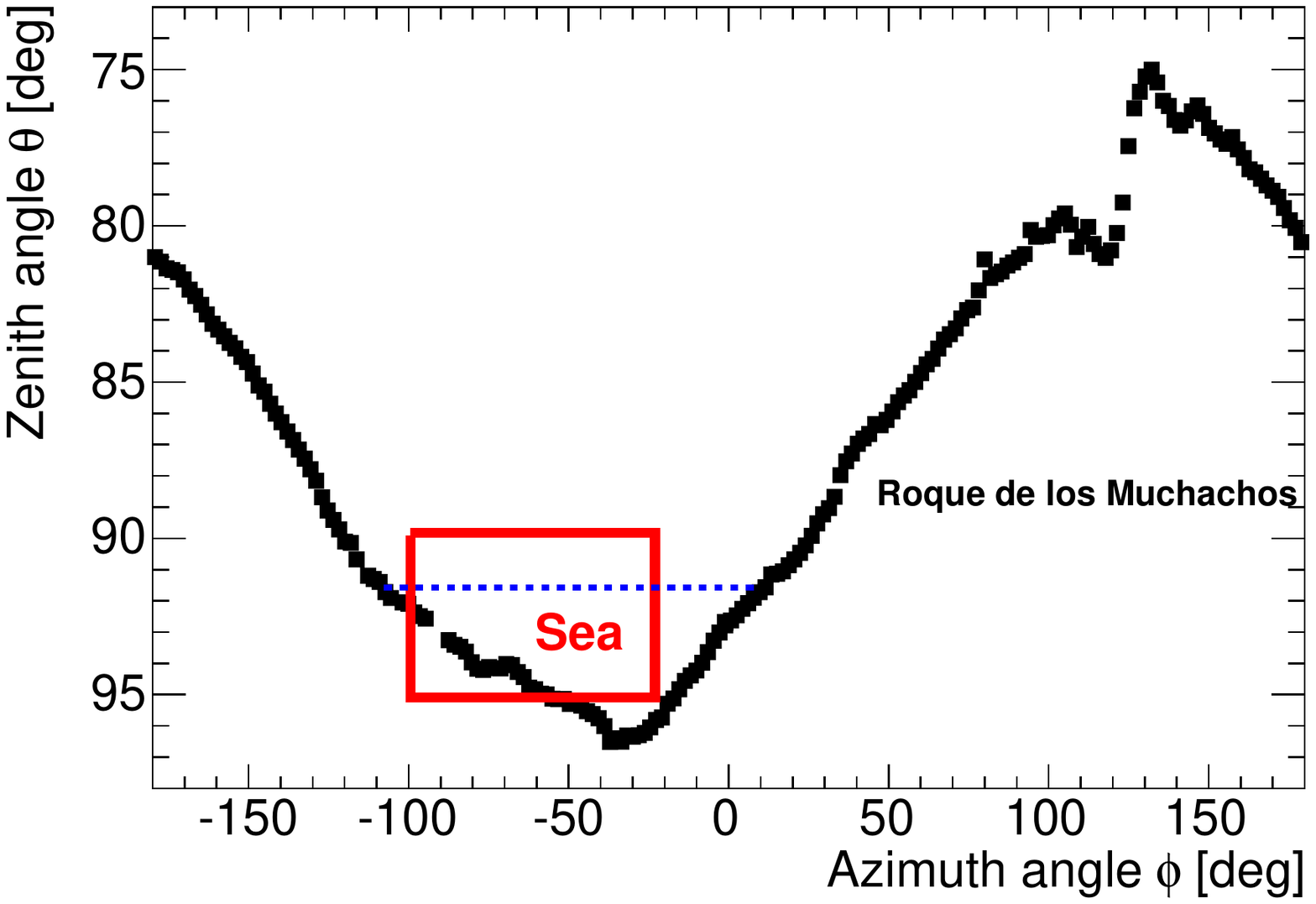}
\caption{Left panel: Illustration of a neutrino converting after distance $x$ into a tau lepton inside the Earth with radius
$R_{\rm Earth}$
and crossing the layer of the Sea towards the MAGIC telescopes. Right panel: the horizon seen from the MAGIC telescopes. The region with azimuth $\phi$ in the range from -20 deg to -$100$ deg  and zenith angle $\theta$ 
from $\sim$90 deg to  $\sim$95 deg can be used to point toward the sea.
The region with azimuth from -20 deg to 0 deg is  excluded, due to shadowing by  the telescopes access towers. }\label{fig::sea}
\label{tab:aa}
\end{figure*}
The detection  of $\nu_{\tau}$  is very important from both the astrophysical and  particle physics point of view. 
It would give new information about the astrophysical $\nu_{\tau}$ flux and  serve 
as 	an additional confirmation of the astrophysical origin of the IceCube high energy diffuse neutrino signal~\cite{icecuflavour,icecuflavour1}. It also 
would shed light on the emission mechanisms at the source, test the fundamental properties of neutrinos over extremely long baselines, and better constrain new physics models which predict significant deviations from equal fractions of all flavors.

\begin{table*}[h]
\caption{Summary of the data taken at very large-zenith angles during MAGIC observations from October 2015  to  March 2017.}\label{tab:a}
\center
\begin{tabular}{lcccc}
\hline
\hline
 & {\it seaOFF } & {\it seaON }   & Roque  & HET   \\
\hline
\hline
Zenith angle $\theta$ ($^{\circ}$)&87.5 & 92.5 & 89.5 & 85-93  \\
\addlinespace
Azimuth  $\phi$ ($^{\circ}$) &-30 & -30 & 170 & -80 - (-75)  \\
\addlinespace
Observation time (h) & 9.2 & 31.5 & 7.5 & 4.5  \\
\hline
\hline
\end{tabular}
\end{table*} 

 Large detectors like  IceCube\footnote{\url{http://icecube.wisc.edu}}, ANTARES\footnote{\url{http://antares.in2p3.fr}}, the Pierre Auger Observatory~\cite{augerneutrino}, the Telescope Array~\footnote{\url{http://www.telescopearray.org}}, 
or radio detectors like ANITA~\cite{anita}, have the capability to detect  neutrino induced showers. Especially if tau
neutrinos interact close to the Earth surface, the so-called Earth's skimming neutrinos~\cite{fargion,feng,bertou}   can produce tau leptons which can emerge from the Earth, decay and produce extended air showers. If the decay vertex of a tau lepton  is close enough to a surface detector, it can be detected and distinguished 
from very inclined proton and nuclei induced showers due to the presence of its electromagnetic component.

Above PeV energies, the Earth becomes opaque to electron and muon neutrinos, while the tau neutrino flux is regenerated through subsequent tau lepton decays to neutrinos. At  high-energies, the tau neutrino interacts in the Earth producing a tau lepton which in turn
decays  into  a $\nu_{\tau}$ with  lower  energy  due  to  its  short  lifetime. The regeneration chain
 $\nu_{\tau} \rightarrow  \tau \rightarrow \nu_{\tau} $ ... continues until the tau lepton reaches the detector. This effect can lead to a significant enhancement of the tau lepton flux of up to about 40\% more than the initial cosmic flux of tau
neutrinos of energies between 1-100 PeV~\cite{jones,reya}.

Tau neutrinos that pass through the earth crust, are the only ones that can produce particle showers that can be detected  by  Imaging Atmospheric Cherenkov Telescopes (IACTs). In the  interaction  all  neutrino types 
loose part of their energy due to charged current and neutral current interactions.
In the case of $\nu_{\rm e}$'s crossing the Earth, an electron is produced which is rapidly brought to rest in matter. In the case of muon neutrinos $\nu_{\mu}$'s, a muon is created which subsequently decays. However, before decaying the muon propagates through  matter, too. As its radiation length in matter is much smaller than  its decay length at our energies of interest, it loses most of its energy. The produced muon can occasionally escape from the Earth, but it  decays only rarely in the atmosphere since the decay length is about $10^8$ times larger than the one of a tau lepton\footnote{The muon decay length is about $6.6 \times 10^6$ km for a muon energy of 1 PeV.}. Energetic tau neutrinos, produce extended air showers  at the decay of the tau-particle, which itself loses much less energy while it travels through matter and which can be efficiently detected by IACTs and discriminated against background. Thus the earth-skimming method is suitable for the detection of tau neutrinos\footnote{The flux of $\mu$ leptons produced  due to charged current interaction of $\nu_{\mu}$ is more than one order of magnitude lower than for $\tau$ leptons \cite{kusenko, hawcneutrino}.}.

 To detect neutrino-induced showers with an IACT system, it needs to be pointed towards the ground,
 e.g. the side of a mountain or  the sea surface~\cite{fargion,Asaoka:2012em,Sasaki:2014,gora:2015,gora:2016}. 

 In this paper we present  limits on the flux of tau neutrinos from the MAGIC  telescopes. MAGIC is  located at the Roque de los Muchachos Observatory  at an altitude of about 2200 m  above sea level (28.8$^{\circ}$ N, 17.9$^{\circ}$ W), in the Canary Island of La Palma (Spain). The observatory consists of two telescopes placed at a distance  of  85 m from one another. The MAGIC telescopes  have a  mirror of 17 m diameter and  a field of view (FOV) of 3.5$^{\circ}$. They have been built to detect cosmic $\gamma$-rays in the energy range from $\sim$50 GeV  to $\sim$50 TeV~\cite{magicperformance}. For this search, MAGIC
was used as a neutrino detector,  in order to  look for air showers induced by tau neutrinos  ($\tau$-induced showers) in the  PeV to EeV energy range. Here we report final results of our preliminary  studies presented in~\cite{icrc2017, epsvhe2017}.

The search of tau neutrinos with MAGIC is performed pointing the telescopes in the direction of $\nu_{\tau}$ which escape the Earth crust and later cross the ocean (see Figure~\ref{fig::sea}, left). 
The telescopes can point up to 6 degrees below the horizontal plane, covering an azimuthal range of 80 degrees (see Figure~\ref{fig::sea}, right). 
The location of MAGIC contains the right distance of the telescope to the average point  of the tau lepton decay vertex. This distance should be at least a few tens of times larger than  the decay  length  of the  tau lepton. At 10/1000 PeV  the tau decay length is   about  $0.5$/$50$ km. If  such a condition is not fulfilled, the induced air-shower is not fully developed, leading to a too small  amount  of produced  Cherenkov light reaching the Cherenkov telescopes.

 In~\cite{upgoing_magic}, the effective area for up-going tau neutrino observations for  MAGIC  was calculated analytically and found to reach $5 \cdot 10^5$~m$^2$ at 100~EeV. An analytical approximation results in tau neutrino effective areas from $\sim$10$^3$~m$^2$ (at 100 TeV) to $6  \times 10^4$ m$^2$ (at 300 PeV) for an observation angle of about  1.5$^{\circ}$ below the horizon\footnote{In\cite{upgoing_magic}  the fluorescence emission above 300 PeV was also  included in the aperture calculation, thus   the  aperture can  reach  about $2 \times 10^5$ m$^2$  at 1 EeV. In this work we do not included fluorescence emission.}, rapidly diminishing with larger inclination. However, the  sensitivity for diffuse neutrinos was found to be  very poor compared to the IceCube or Pierre Auger experiments due to the limited FOV, and the shorter observation times with MAGIC.

%
%

In the case the telescopes are pointed to flaring or  disrupting point sources such as gamma ray bursts (GRBs) or active galactic nuclei (AGNs), one can expect to observe a signal from neutrinos.  Indeed, it was shown  by   the  Ashra (All-sky Survey High Resolution Air-shower detector) team \cite{Asaoka:2012em} and by~\cite{upgoing_magic}, that Cherenkov telescopes  can be sensitive to close-by GRBs ($z < 0.1$). It is also known  that a large amount of rock 
surrounding the site, like mountains,  can lead to a significant enhancement of the tau lepton flux, see for example~\cite{gora:2015}. However,  in the case of the MAGIC 
site, the mountain  is too close to the  telescopes, and the possible $\tau$-leptons emerging from the mountain would not have sufficient time to create 
the electromagnetic showers before reaching  the telescopes.

 It is worth to mention that this kind of observations can be performed during the presence of high clouds above the detector.
In such a case, the regular MAGIC gamma-ray observations  are not possible, but  such conditions allow to perform horizontal observations for tau neutrinos. The amount of observation time varies from one to another MAGIC observation season, but amounts to about 100 hours per year~\cite{frac}. 

\begin{figure*}[ht!]
 \centering
 \noindent
 \includegraphics [width=0.27\textwidth]{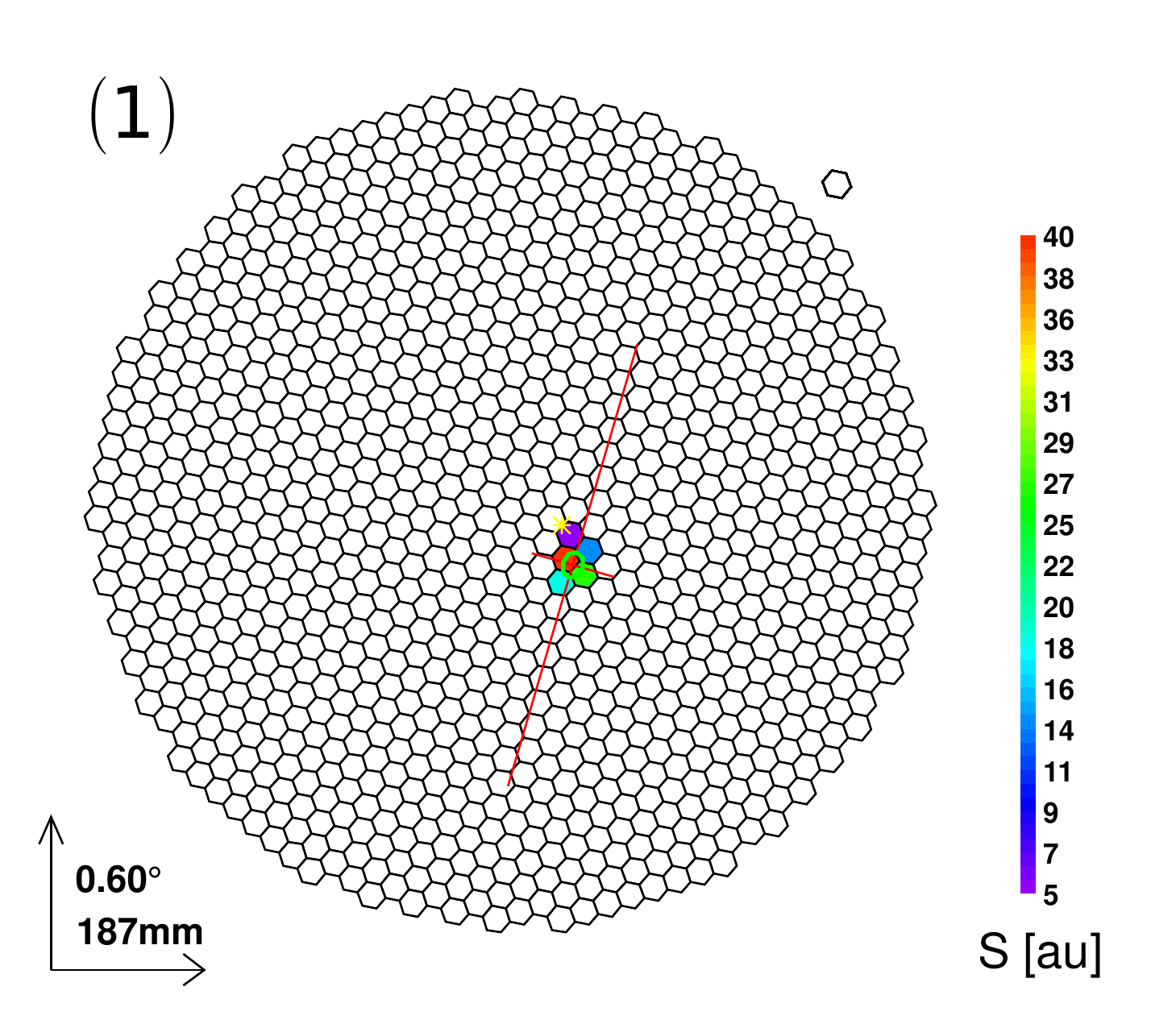}
\includegraphics [width=0.27\textwidth]{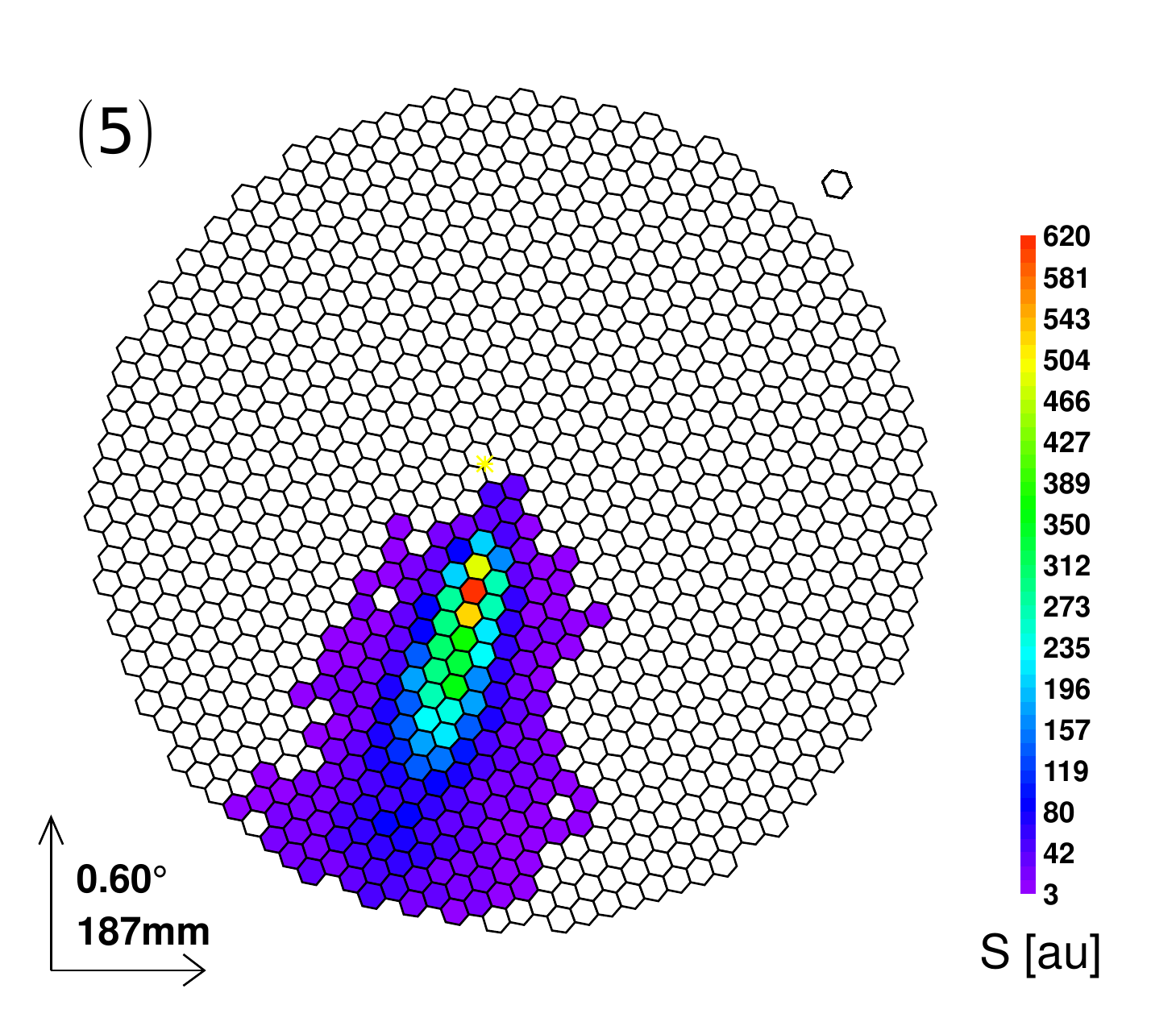}
 \caption{\small Examples of simulated events triggered stereoscopically and shower images for 1 PeV protons/muon injected at the top of the atmosphere under a  zenith angle of $\theta\simeq 86^{\circ}$ (background expectation),
 as seen by one of  the MAGIC telescopes. The first interaction point is at a vertical depth, defined  as
the starting altitude in g/$cm^2$ of the mass overburden of the atmosphere,  below 50 g/cm$^2$ and
a distance between the detector and the proton interaction point  of about  800 km. Example of  (1)  CR showers at lower
energies (see text),  (2)  a single  1 PeV muon as the primary particle, injected at the top of the atmosphere,  interacting via radiative processes and producing a bright gamma-like image on the camera.}\label{fig::backimages}
\end{figure*}

The structure of this paper is the following: Section 2 describes  the recent MAGIC observation at very large zenith angles ( $>85^{\circ}$) and presents the Monte Carlo (MC) simulation chain. In  Section 3, we study the properties of shower images on the camera focal plane from $\tau$-induced showers,  and we show that MAGIC 
 can discriminate $\tau$-induced showers from the background of large-zenith angle  cosmic-ray (CR)  induced showers. In Section 4 and 5
details of the acceptance calculations are presented, together with the expected event rates and the MAGIC sensitivity for tau neutrinos. 
Finally,  in Section 6, a  short summary is given. 

\section{MAGIC observations and Monte Carlo simulations}
The MAGIC telescopes have collected approximately 30 hours of data at very large zenith angles 
($\theta = 92.5^{\circ}$) in the direction of Sea, referred to as {\it seaON}. Events from slightly above the sea ({\it seaOFF}), towards the Roque de los Muchachos mountain or pointing to the Highest Energy Track (HET)~\cite{het} event from IceCube\footnote{Up to now, the HET is the highest energy neutrino event seen  in the IceCube data,  
with a pointing  accuracy  of about 0.27$^{\circ}$ (median), thus particularly interesting for IACTs, given their FOV of a few degrees.} have also been obtained. Details about all event samples are shown in Table~\ref{tab:aa}. 
A  significant  amount of data, about 91\%, was accumulated during the presence of  high clouds  at the MAGIC site. 

%
%

 The  rate  of events seen by both MAGIC telescopes the so-called  stereo event rate, in case of   {\it seaOFF } data, 
 due to increasing attenuation of the  Cherenkov light as a function of the  zenith angle, is about 27 times  larger   ($\sim$ 4.6 Hz) than  for {\it seaON } ($\sim$0.17 Hz) observations. Thus, {\it seaOFF } observations provide  a high-statistics background estimate for  the  {\it seaON } data.   In principle, the expected  {\it seaON } background should be almost zero. However, due to the specific location of MAGIC, the curvature of Earth and  the fact that the MAGIC telescopes can point up to about 6 degrees below horizon, the contribution of cosmic rays is not exactly zero. This also means that in the case of MAGIC telescopes with  FOV of $3^{\circ}\times3^{\circ}$  the expected signal of tau neutrinos on the camera is not uniform as a function of zenith angle. 
 The camera sees not only the earth crust but also part of the atmosphere or the sea. In principle, pointing to the
  larger inclinations ($> 95$ degrees) should decrease the fraction of atmosphere seen  by the  camera  and significantly reduce the background level,  but  in such a case  the expected flux of tau neutrinos decreases as well. In \cite{upgoing_magic} it was  found that the maximum  effective tau neutrino area on the camera  is  at 91.6 degrees, and  reduces  almost to   zero   at 90.5 degrees, and is more than one order of magnitude smaller, in comparison with its  maximum value,  at 93.5 deg.  It was also found that the maximum effective neutrino  area integrated  over the MAGIC FOV is for  pointing at 92.5 degrees. This allows to catch the most sensitive part in the outskirts of the camera and add the additional contribution of the effective area above $ 92.5^{\circ}$. Thus, the {\it seaON  } data were taken with a zenith angle of $\theta=92.5^{\circ}$.

The software tool used in the present paper to simulate the signatures expected from neutrino-induced showers 
by MAGIC is based on the code ANIS  (All Neutrino Interaction Simulation)~\cite{anis} with some extension described in~\cite{goraanis}.
\begin{table*}[h]  
\small                                                                                                                                                                   
  \caption{                                                                                                                                                                          
    \label{tab::decay-modes}                                                                                                                                                         
     Tau lepton decay channels implemented in ANIS simulations. The Table is                                                                                                      
    taken from \cite{fargion}.} 
\center                                                                                                                                               
  \begin{tabular}{ccccc}                                                                                                                                                             
    \hline                                                                                                                                                                           
    \hline                                                                                                                                                                           
 Decay        &    Secondaries         &  Probability  & Air-shower  \\                                                                                                              
             &    &    &   &  \\                                                                                                                                                     
\hline 
\hline                                                                                                                                                                              
$\tau \rightarrow \mu^{-}\bar{\nu}_{\mu} \nu_{\tau} $&  $\mu^{-}$  &  17.4\%                                                                                                         
& weak showers   &  \\                                                                                                                                                               
$\tau \rightarrow  e^{-}\bar{\nu}_{e} \nu_{\tau} $&  $e^{-}$  &  17.8\%  &1                                                                                                         
Electromagnetic  &  \\                                                                                                                                                               
$\tau \rightarrow \pi^{-} \nu_{\tau} $&  $\pi^{-}$  &  11.8\%  & 1 Hadronic &                                                                                                        
\\                                                                                                                                                                                   
$\tau \rightarrow \pi^{-}\pi^{0} \nu_{\tau} $&  $\pi^{-}$, $\pi^{0}\rightarrow                                                                                                       
2\gamma$  &  25.8\%  &1 Hadronic, 2 Electromagnetic  &  \\                                                                                                                           
$\tau \rightarrow \pi^{-}2\pi^{0} \nu_{\tau} $&  $\pi^{-}$,                                                                                                                          
$2\pi^{0}\rightarrow 4\gamma$  &  10.79\%  &1 Hadronic, 4 Electromagnetic  &                                                                                                         
\\                                                                                                                                                                                   
$\tau \rightarrow \pi^{-}3\pi^{0} \nu_{\tau} $&  $\pi^{-}$,                                                                                                                          
$3\pi^{0}\rightarrow 6\gamma$  &  1.23\%  &1 Hadronic, 6 Electromagnetic  &                                                                                                          
\\                                                                                                                                                                                   
$\tau \rightarrow \pi^{-}\pi^{-} \pi^{+}\nu_{\tau} $&  $2\pi^{-}$,$\pi^{+}$  &                                                                                                       
10\%  &3 Hadronic &  \\                                                                                                                                                              
$\tau \rightarrow \pi^{-}\pi^{+} \pi^{-} \pi^{0} \nu_{\tau} $&                                                                                                                       
$2\pi^{-}$,$\pi^{+}$,$\pi^{0}\rightarrow2\gamma$  &  5.18\%  &3 Hadronic, 2                                                                                                          
Electromagnetic  &  \\                                                                                                                                                               
\hline                                                                                                                                                                               
\hline                                                                                                                                                                               
\end{tabular}                                                                                                                                                                        
\end{table*}                                                                                                                                                                                                                                                                                                                                                            
CORSIKA version 6.99~\cite{corsika}, with  activated CERENKOV, CURVED-EARTH, TAULEP and THIN options,   was used to  simulate the shower development of $\tau$-induced showers and its Cherenkov light production. Tau lepton decays have been simulated with PYTHIA version 6.4 \cite{pythia}. The output of the CORSIKA simulations are then passed to a simulation of the atmospheric extinction and the MAGIC Telescope response~\cite{mars}. Due to technical difficulties, only showers simulated between 86 degrees and 90 degrees have been used to study the properties of upward-going tau neutrinos. This limitation is not a problem, as the response of IACTs to Cherenkov light from showers of the same energy and equal column depth depends only slightly on the zenith angle~\cite{gora:2016}.
In our MC, we also performed simulations for inclined showers induced by CRs. This gives the possibility to compare  simulated images on the MAGIC camera plane with images from real data.

In the studied energy range, from 1 PeV to 1 EeV,  the main background for IACTs are proton induced showers. The background from photon and electron induced showers can be expected to be considerably smaller,  especially beyond the spectral break in the electron spectrum at about 1~TeV~\cite{hesseleectron}.

In the case of showers  induced by CRs and observed at large zenith angles,  the hadronic and electromagnetic component of extensive air showers (EAS)
is almost  absorbed~\cite{innes} because  of the deep horizontal column depth of about $\simeq 10^{4}$ up to $5\times10^{4}$ g cm$^{-2}$ in such directions.  
In the case of  CR showers of lower energy (from tens of GeV up to PeV) only a  few pixels in the camera
 are triggered, yielding  dimmer and smaller images (see  Figure~\ref{fig::backimages}  left).
 The larger shower images come from  high energetic  muons (from tens to hundred of GeV) produced at the first stages of shower development or muon bundles
 from later stages.
Penetrating muons from  bundles can  decay  into electrons not far from the telescopes, which  induce small air-showers  producing detectable flashes  of  Cherenkov light. 
This produces  different image topologies in the camera, e.g. a few clusters  of triggered pixels,  as is  shown in~\cite{icrc2017}.
The estimated event rate  for such showers in the case of a MAGIC-like detector  is at the  level of about one event per two minutes for CR showers of energies above $E_{\mbox{\small CR}}> 60$ PeV~\cite{fargion:0511597}.

High energy muons (E $>$ 1 TeV) at large zenith angles have an interaction length, via $e^+e^-$-pair production, bremsstrahlung, and photo-nuclear scattering, comparable to the depth of the atmosphere. 
In these events produced by interacting muons via radiative processes, additional electromagnetic sub-showers are induced~\cite{kiraly,kiraly2}. If these sub-showers are induced close to the detector, it can lead to a strong flash of Cherenkov light and a bright image in the camera (see Figure~\ref{fig::backimages}  right). All classes of simulated events shown in Figure~\ref{fig::backimages} and  presented also in~\cite{icrc2017} have been observed in the data taken with the MAGIC telescopes at large zenith.
{\center
\begin{figure*}{}
\begin{minipage}[b]{0.49\textwidth}
\includegraphics [width=0.95\textwidth,height=8.0cm]{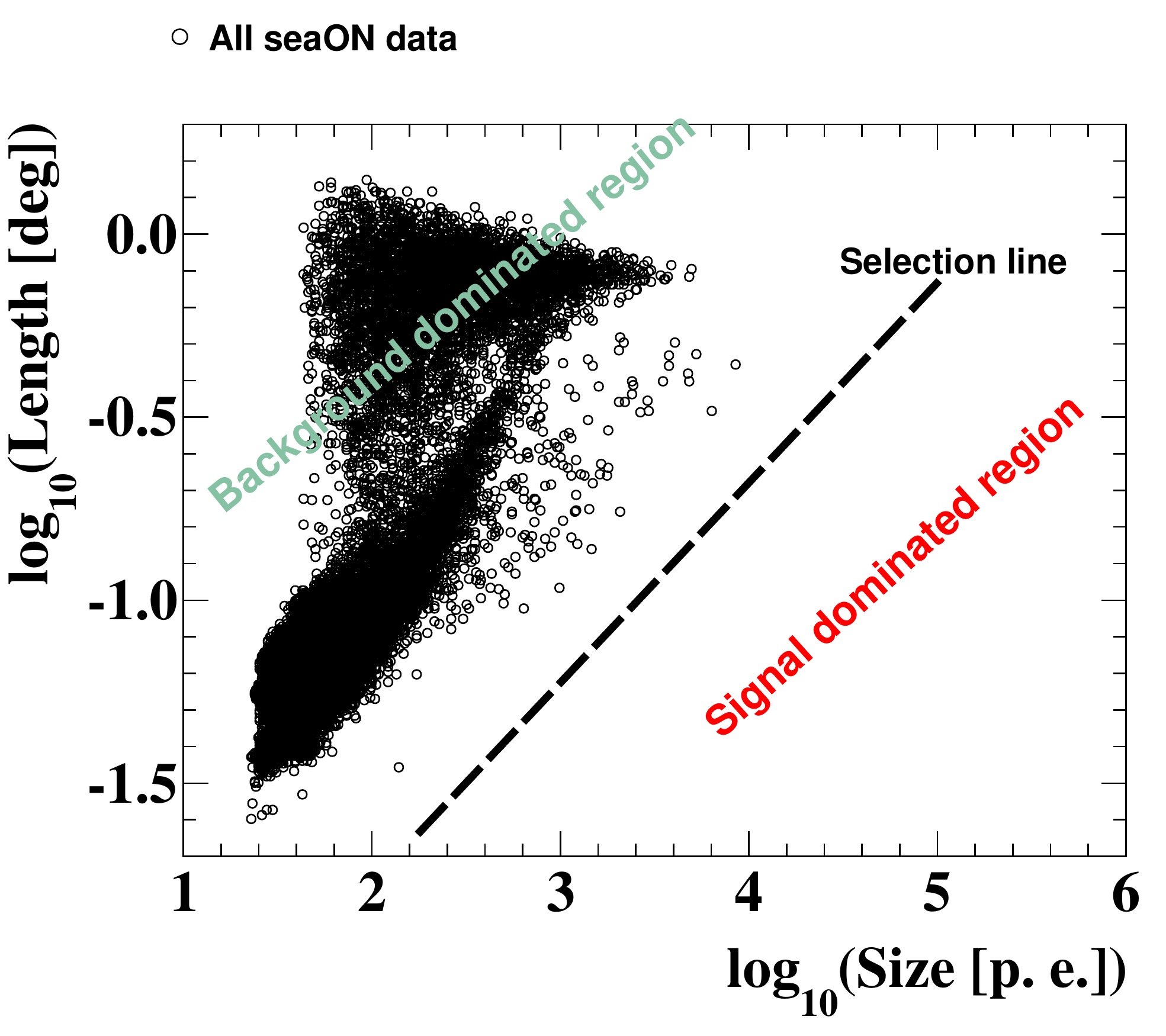}
 \end{minipage}
\begin{minipage}[b]{0.49\textwidth}
\includegraphics [width=\textwidth,height=7.5cm]{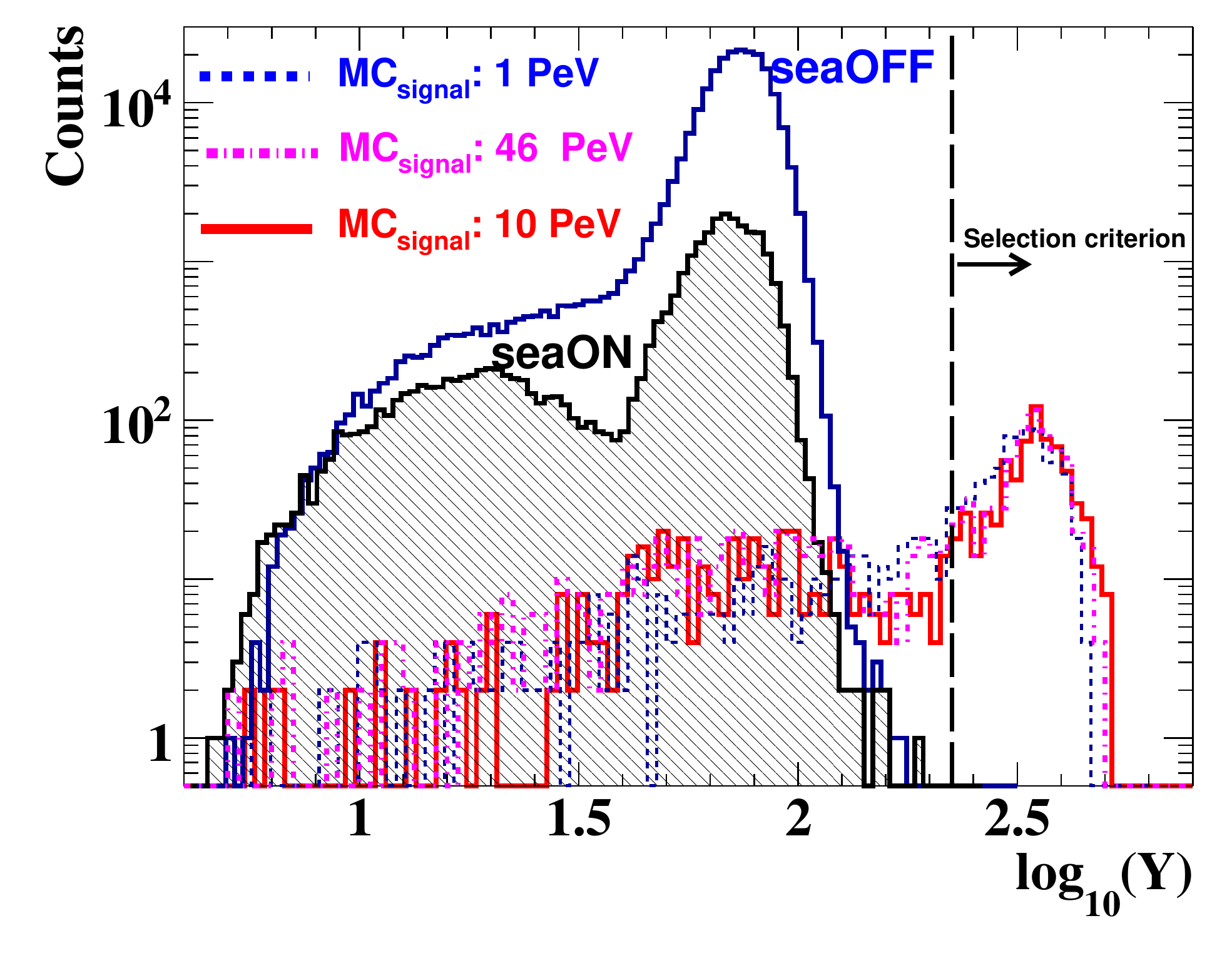}
\end{minipage} 
\caption{ \small Left panel:   2D distribution of {\it seaON} measured events in the Hillas \textit{Length} versus \textit{Size} paramater plane. In the signal dominated region, no events remain after selection. Right panel: Distribution of events in the direction perpendicular to the selection line, defined by  Y-axis (see text), for {\it seaON}, {\it seaOFF} and signal MC events. In the region $\log_{10}(Y)>2.35$, no neutrino candidates remain. The selection criterion keeps about $\sim$40\% of  MC signal events.} \label{fig::dist2}
\end{figure*}
}

The expected signature of tau leptons in the camera depends on the different tau decay channels. Tau leptons decay mainly to hadrons, pions and kaons, and in ~35\% of the cases to electrons and muons (see  Table~\ref{tab::decay-modes}).  In~\cite{icrc2017}    we show  
shower images from our MC signal simulations for a  1 PeV tau lepton decaying into an electron, pion or  muon close to the detector,  i.e. for a  typical detector-to-shower  
distance  of about  50 km. The showers produced by tau leptons have in general  larger size and contain many more photons compared to the ones of the background coming from protons. 
This is not only due to the closer distance to the detector, but also
 because   the  tau lepton mostly decays into electrons and pions, which induce electromagnetic sub-showers producing a large amount of Cherenkov light.
Electromagnetic showers  or a superposition of electromagnetic sub-showers come from decays of neutral pions, while  hadronic sub-showers come from   interaction and decay of charged pions. Tau leptons can also decay into muons, see Table~\ref{tab::decay-modes}. At energies of $>$ 1 PeV, the muon has a large interaction length of a few thousand kilometers in air, so the muon mainly interacts with the atmosphere through secondary bremmstrahlung processes ~\cite{muonshower}. This makes the muon ring image hardly visible in the camera~\cite{icrc2017} .

\section{Discrimination of $\tau$-induced showers }
Simulated events of $\tau$-induced showers are calibrated following the same procedure as real data. The number of photoelectrons per camera pixel are extracted using a sliding window algorithm~\cite{magicperformance}. In order to remove the pixels which are most likely due to the night sky background, an image cleaning procedure is carried out~\cite{magicperformance}. The cleaned camera images are characterized by a set of image parameters which were introduced by Hillas~\cite{hillas}.

 These parameters provide a geometrical description of the images of showers and are used to infer the energy of the primary particle, its arrival direction, and to distinguish between $\gamma-$ray and hadron induced showers. A typical spatial distribution of Cherenkov photons on the camera can be parameterized as an ellipse. The rms spread of Cherenkov light along the major/minor axis of image is known as the {\it Length}/{\it Width} of an image. The {\it Length} and the {\it Width} parameter are a measure of the lateral and the vertical development of the shower. The  {\it Size} parameter measures the total amount of detected  light (in p.e.)  in all camera pixels, and is correlated
with the primary energy of the shower. In the following we  study  these parameters also  for the case of  deep i.e. with the first interaction point deep in the atmosphere,  $\tau$-induced simulated showers and compare the corresponding distributions with  data. 

\begin{figure*}[ht!]
 \centering
\includegraphics [width=0.48\textwidth,height=7cm]{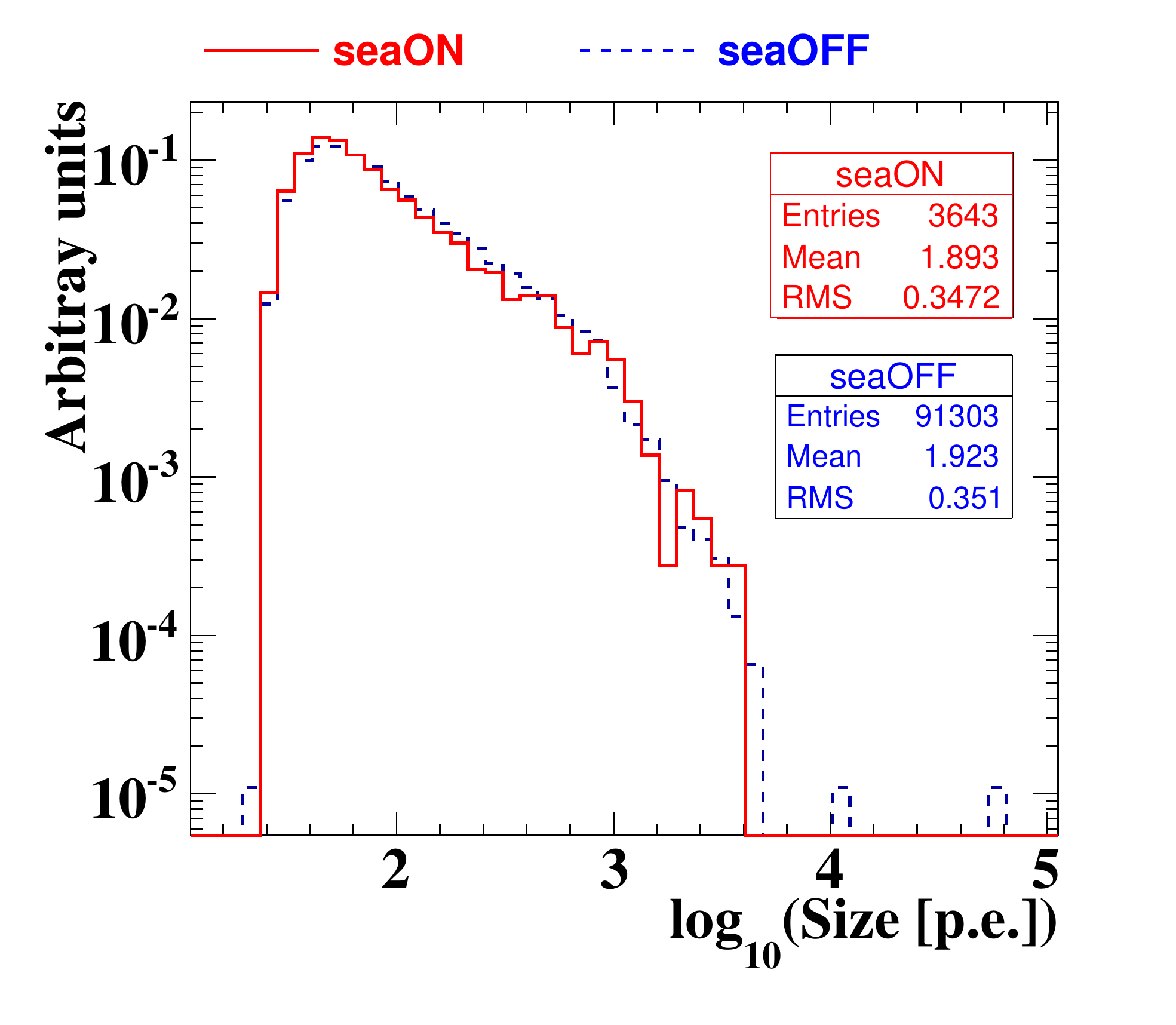}
\includegraphics [width=0.48\textwidth,height=7cm ]{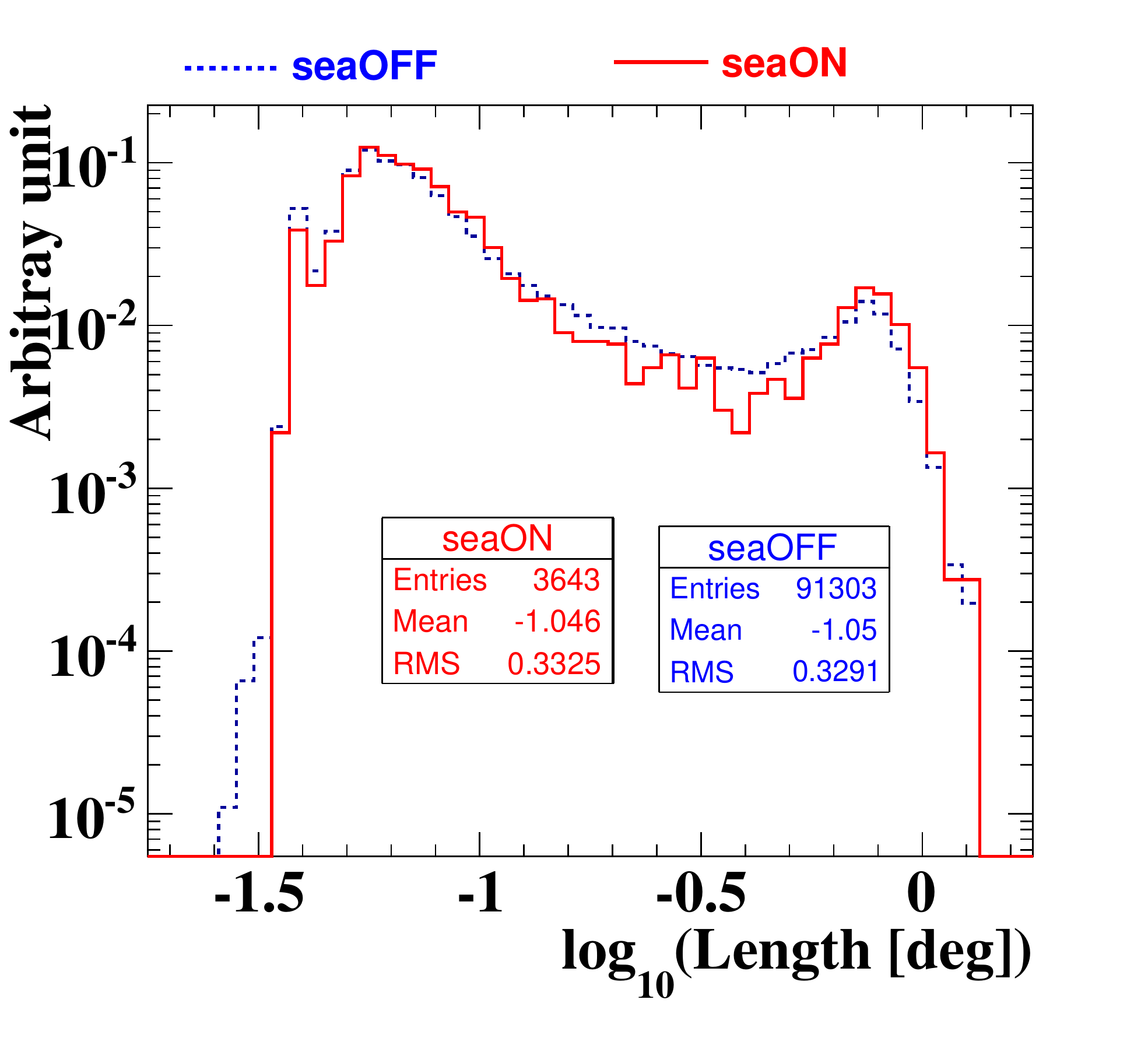} 
 \caption{\small  Normalized distribution of the Hillas parameters \textit{Size} (left panel) and  \textit{Length} (right panel) for {\it seaON } and {\it seaOFF } data.}\label{zenithbins}
\end{figure*} 

The MAGIC telescopes took data at zenith angles of 92.5 degrees ({\it  seaON  }) and 87.5 degrees ({\it seaOFF }). In the {\it seaOFF } data, a negligible signal of neutrinos events is expected. This is due to the fact, that the relative rate of neutrino interactions is lower for {\it seaOFF } because of the Earth skimming channel enhances the rate, although this enhancement may not be so high at 1 PeV.
The {\it seaOFF } data is used to estimate the background contribution in the {\it seaON } data and also to construct the selection criterion to identify tau-neutrino showers, see~\cite{epsvhe2017} for  a detailed description.

%

Here, in Figure~\ref{fig::dist2} (left)  we show a 2D distriution of {\it Size} versus {\it Length} parameter for all  {\it seaON} data, but 
in  Figure~\ref{fig::dist2} (right) the distribution of the {\it seaON}, {\it seaOFF} and signal MC events in the direction perpendicular to the  selection  line.
The  Y coordinate  was obtained from  the following formula: $\log_{10}(Y)=\log_{10}(\textit{Size}[\mathrm{p.e.}])*\cos(\alpha)-\log_{10}( \textit{Length}[\mathrm{deg}])*\sin(\alpha)$, where $\alpha=63.435^{\circ}$. In this representation  the selection line  shown in Figure~\ref{fig::dist2} (left)   transforms to
 a single cut value  given by: $\log_{10}(Y)=2.35$.  For showers with   impact distance smaller than 0.3 km we use the following
  selection cut:  $\log_{10}(Y)>2.35$,  while  for showers with  larger impact distance (0.3 - 1.3 km) a slightly  relaxed cut  was used i.e.  $\log_{10}(Y)>2.10$.  The impact distance  is defined as the distance between the shower axis and the telescope axis, see  also Figure~\ref{sketch}. 
  The distribution of impact distances in CORSIKA simulations can be set by using the option CSCAT~\cite{corsika}. Their reconstruction procedure has been described in~\cite{magicperformance}.

 In any case we did not find  any neutrino candidate, if this  selection criterion is applied  to  all  {\it seaON } data, see Figure~\ref{fig::dist2}. The  selection criterion was  optimized by maximization of MC signal efficiency, which finally  reached  a  level 
of  about 40\%, and minimization of the  background contribution. The  selection cut  was   placed  in the Y range where the signal distribution starts to be flat, in order to avoid a significant drop of the signal efficiency for larger values of the Y cut parameter.

As can be seen from the  Figure~\ref{fig::dist2} (right)  the  MC  signal distributions  look  rather similar, as a consequence of  the  small  dependence  of the Hillas parameters on  the primary energy of the  tau lepton, as shown in~\cite{icrc2017}.
 
It is  also worth noting that we see  the universal character of the Hillas parameters at large zeniths. This is seen in MC
simulations~\cite{gora:2016} but also in MAGIC data taken  during periods of good weather conditions. As an example, in  Figure~\ref{zenithbins}  the normalized distribution of Hillas parameters  for  data taken in the  {\it seaOFF } ($\sim$5.5 hrs, 87.5$^\circ$ zenith angle)  and {\it seaON  } ($\sim$6 hrs, 92.5$^\circ$ zenith angle) direction are shown. The data were taken during one night, under similar weather conditions.  As we can see, the {\it Size} distribution  for {\it seaOFF } data  is similar  compared to {\it  seaON } data. A similar behavior is  seen for the {\it Length} parameter. In the  case  when we merge  larger  time periods, the shape of the  Hillas  {\it Size} and  {\it Length} parameters may be slightly different from one night to another breaking this universality behavior. This is  due to different weather conditions and thus different   attenuation of Cherenkov light. The effect influence  a small fraction of  low energetic events with $\log_{10}(Y) <1.6$ (see as an example  Figure~\ref{fig::dist2}, right).  Figure~\ref{zenithbins}  and  Figure~\ref{fig::dist2}  support  our previous assumption when performing  MC simulations  only at large zenith angles, in order to study the response of MAGIC  for upward-going $\tau$-induced showers.

\section{Event rate estimation }
In this section, we discuss  first different  models  for neutrino production, then 
show details on the calculations of the acceptance of the MAGIC telescopes, and finally we present the expected
 sensitivity for $\tau$-induced showers  in MAGIC.  

\subsection{Astrophysical target flux}
 
The detection of ultra-high energy neutrinos, with energies in the PeV range or above, is a topic of great interest in modern astrophysics.
The importance comes from the fact that these neutrinos point back to the most energetic particle accelerators in the Universe, and provide information about 
their underlying acceleration mechanisms. Neutrinos in the PeV range and above are suspected to be produced by AGNs~\cite{atoyan,neronow,mucke},  GRBs~\cite{grbneutrino}  
or   by the interactions of ultra-high-energy cosmic rays with low energy photon backgrounds, 
such as the cosmic microwave background  and the extragalactic background light~\cite{gzkneutrinos,gzkneutrinos1,gzkneutrinos2,gzkneutrinos3, gzkneutrinos4}.

The IceCube Collaboration has  recently  reported the first observation of a cosmic diffuse neutrino flux (all-flavours) in the 100 TeV to PeV range,  which can be described by the  power law~\cite{icecuflavour1,icrc2017hese}:
\begin{equation} 
\frac{d\Phi_{\nu}(E_{\nu})}{dE_{\nu}} = \phi \times \left( \frac{E_{\nu}}{100 \; \mbox{TeV}} \right)^{-\gamma}\label{neutrinoflux}
\end{equation}
where  $E_{\nu}$ is  energy of neutrino.

The best-fit power law corresponds to a normalization 
$\phi = 6.7^{+1.1}_{-1.2}\times10^{-18} \mbox{ GeV$^{-1}$ cm$^{-2}$ sr$^{-1}$ s$^{-1}$} $ and $\gamma = 2.50\pm 0.09$. A somewhat harder
slope $\gamma =2.2\pm0.2$ is found in the muon neutrino signal for upgoing muon neutrinos detected above $\sim 100$ TeV~\cite{muonicecubes}.
The spectrum of the neutrino signal is for the  all-flavour high-energy starting events with neutrino vertex contained in the detector volume. Individual neutrino sources, however, could not be identified up to now. While many
sources of astrophysical  origin have been suggested to be responsible for the IceCube signal, like for example star-forming and/or starburst galaxies~\cite{starburst,starburst1,starburst2,starburst3,starburst4} there is not yet enough information to narrow down the possibilities to any particular source or source class. For a review on possible source candidates see~\cite{Anchordoqui,icecuflavour,icecuflavour1}. 

\subsection{Expected neutrino flux from GRBs}
 GRBs are short gamma-ray flashes lasting from fractions of a second to tens of minutes in most cases. During their prompt emission they are the brightest sources in the Universe. GRBs reach  an isotropic-equivalent energy of up to $10^{54}$ ergs, and are likely powered by the core-collapse of a very massive star or the merger of two compact objects. The central engine produces highly relativistic collimated jets, which are predicted to host internal shocks, where particles are efficiently accelerated to high energies. In hadronic scenarios accelerated protons interact with ambient synchrotron photons and produce high-energy neutrinos  in the PeV-EeV range~\cite{grbneutrino,wintergrb}. The neutrino emission is expected to be collimated and in temporal coincidence with the prompt gamma-ray emission. However, recent results from the IceCube Collaboration~\cite{grbicecube} strongly disfavor classical GRBs as sources of the highest energy cosmic rays and neutrinos.  
Only more complex models assuming multiple emission regions~ \cite{llgrb,wintergrb} can predict a neutrino flux at the level of the IceCube neutrino  astrophysical signal.  
As it is shown in~\cite{llgrb}, a class of interesting objects are the chocked jet GRBs proposed as a way to model and unify the Low Luminosity GRBs  and hypernovae. The basic concept is that a supernova  explosion takes place in a dense surrounding  medium and the emerging jet is a) completely choked, b) partially choked, with a shock front emerging or c) emerging without obstacles. In case a) we expect only a prompt (duration 10$^{1.5}$ s) neutrino emission. In case b) we expect neutrinos from the chocked jet and a delayed (10$^{2}$ -10$^{3}$ s) gamma-ray flare from the shock front, with duration of $> $10$^{3}$ s. Case c) predicts prompt and simultaneous neutrino and gamma-ray emission of 10$^{3.5}$ s duration.

\subsection{Expected neutrino flux from AGNs}\label{agnflux}
For AGNs the probability of discovering extraterrestrial neutrino point sources varies with the supposed  phenomenology
of the accelerators and of their emission mechanisms at high energies~\cite{atoyan,neronow,mucke}. Neutrino emission might be possible for sources where charged and neutral mesons 
are produced simultaneously from hadronic $p-p$ or $p-\gamma$ interactions. These hadronic processes may be present in  variable extragalactic objects such 
as BL Lacs or flat-spectrum radio quasars (FSRQs), as well as in Galactic systems like microquasars and magnetars. 
Blazars, a subset of radio-loud active galactic nuclei with relativistic jets pointing towards the Earth, can significantly contribute to the diffuse (extragalactic) 
$\gamma$-ray background~\cite{ajello}.  If these $\gamma$-rays originate from proton interactions, the energy budget will be 	 sufficient to account 
for the intensity of the IceCube neutrino flux~\cite{starburst2}. However, a recent stacking analysis using IceCube data suggest that AGN blazars contribute at most 27\% of the observed neutrino intensity~\cite{glusen2016}. A recent review  of a possible  correlation between gamma-rays and PeV neutrinos from blazars can be found in~\cite{winter}.
\begin{figure}                                                                                                             
  \begin{center}                                                                                                              
    \includegraphics[width= \columnwidth,height=7cm]{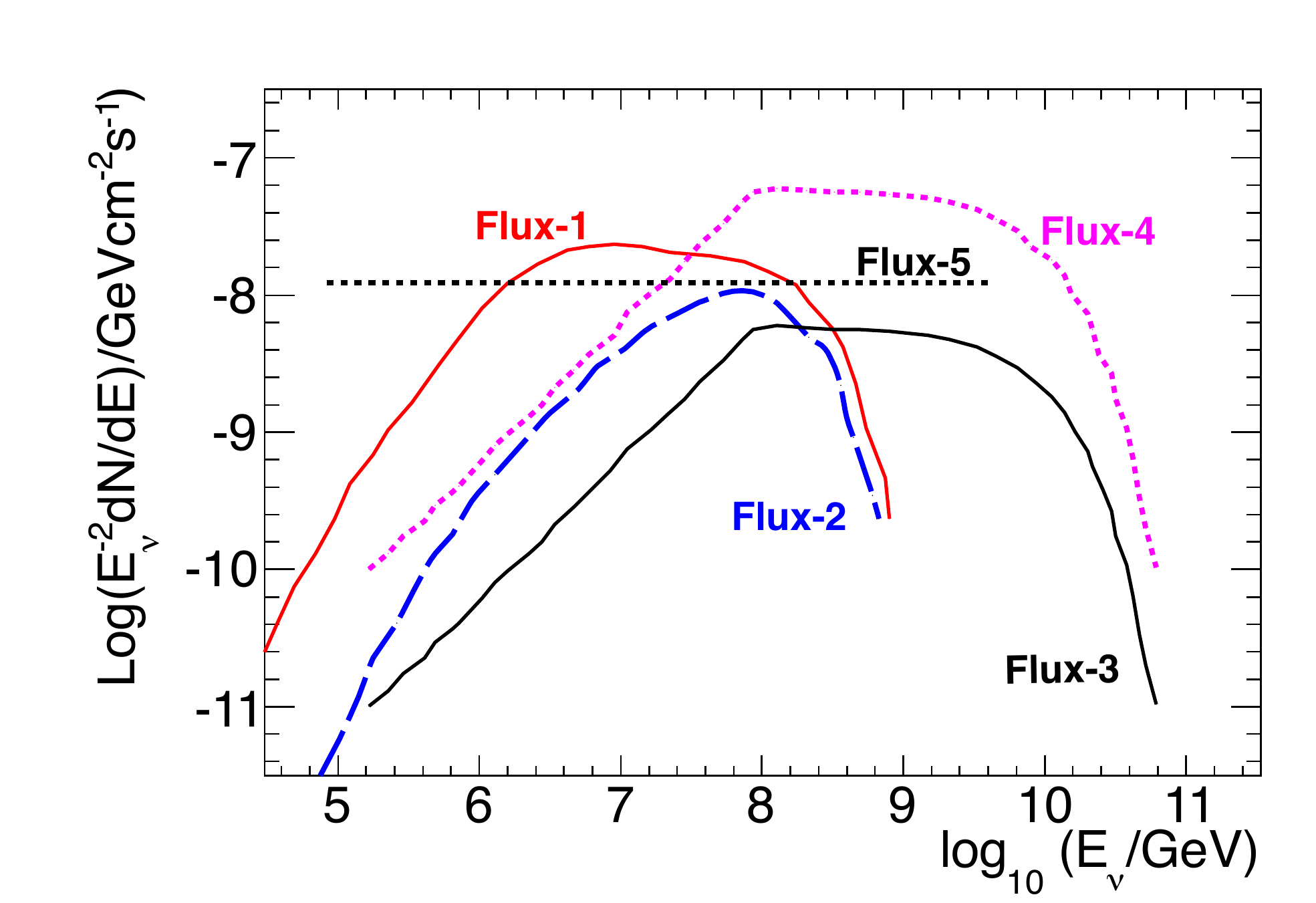}
  \end{center}                                                                                                               
  \caption{\label{fig::spectrum2} Compilation of the neutrino flux predictions from photo-hadronic interactions in AGNs for the $\gamma$-ray flare of 3C 279~\cite{2009IJMPD} (Flux-1 and Flux 2), PKS~2155-304~\cite{Becker2011269}(Flux-3  and Flux-4) and 3C~279~\cite{PhysRevLett.87.221102} (Flux-5). } 
  \end{figure}

Nevertheless, flaring AGNs  can provide a boosted flux of neutrinos  which in some cases could  be at the  level of the IceCube neutrino signal. In this paper we consider predictions for a sample  of generic neutrino fluxes, 
from photo-hadronic interactions in the case of a few powerful  AGNs flares,  as shown in  Figure~\ref{fig::spectrum2}. Flux-1 and Flux-2 are calculations for the Feb 23, 2006 $\gamma$-ray flare of 3C 279~\cite{2009IJMPD}. Flux-3 and Flux-4 represent predictions for PKS~2155-304 in low-state and high-state, respectively~\cite{Becker2011269}. Flux-5 corresponds to a theoretical prediction for 3C~279 calculated in~\cite{PhysRevLett.87.221102}.

\begin{figure}[ht!]
 \centering
 \includegraphics [width=0.49\textwidth,height=6cm]{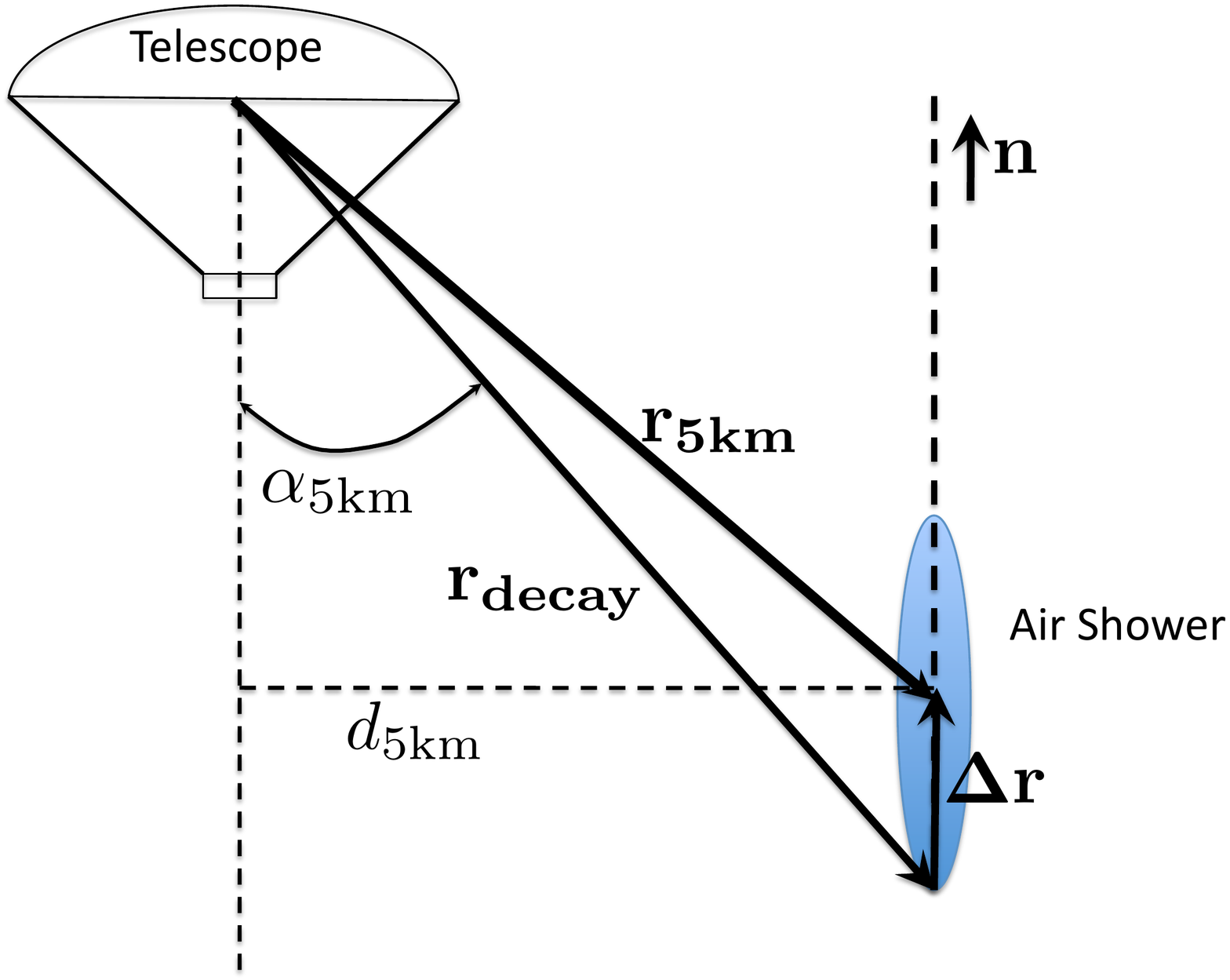}
 \caption{\small Sketch illustrating the different selections cuts performed in the analysis, to guarantee that the estimated position of the shower maximum of simulated events used in the acceptance calculation is in the FOV of MAGIC (see Section~\ref{acceptnace:chapter} for details). The  subscript of 5 km corresponds to estimate  position of the shower maximum, which  is   for  shower energies relevant in this analysis i.e. 1 PeV - 3 EeV  is  reached approximately  after  600 g/cm$^2$ on average, which in the lower part of the atmosphere  translates to  the distance   of about 5 km.}\label{sketch}
\vspace{-0.25cm}
\end{figure} 
\subsection{Background estimation}
\begin{figure*}[ht!]
 \centering
 \includegraphics [width=0.49\textwidth,height=6cm]{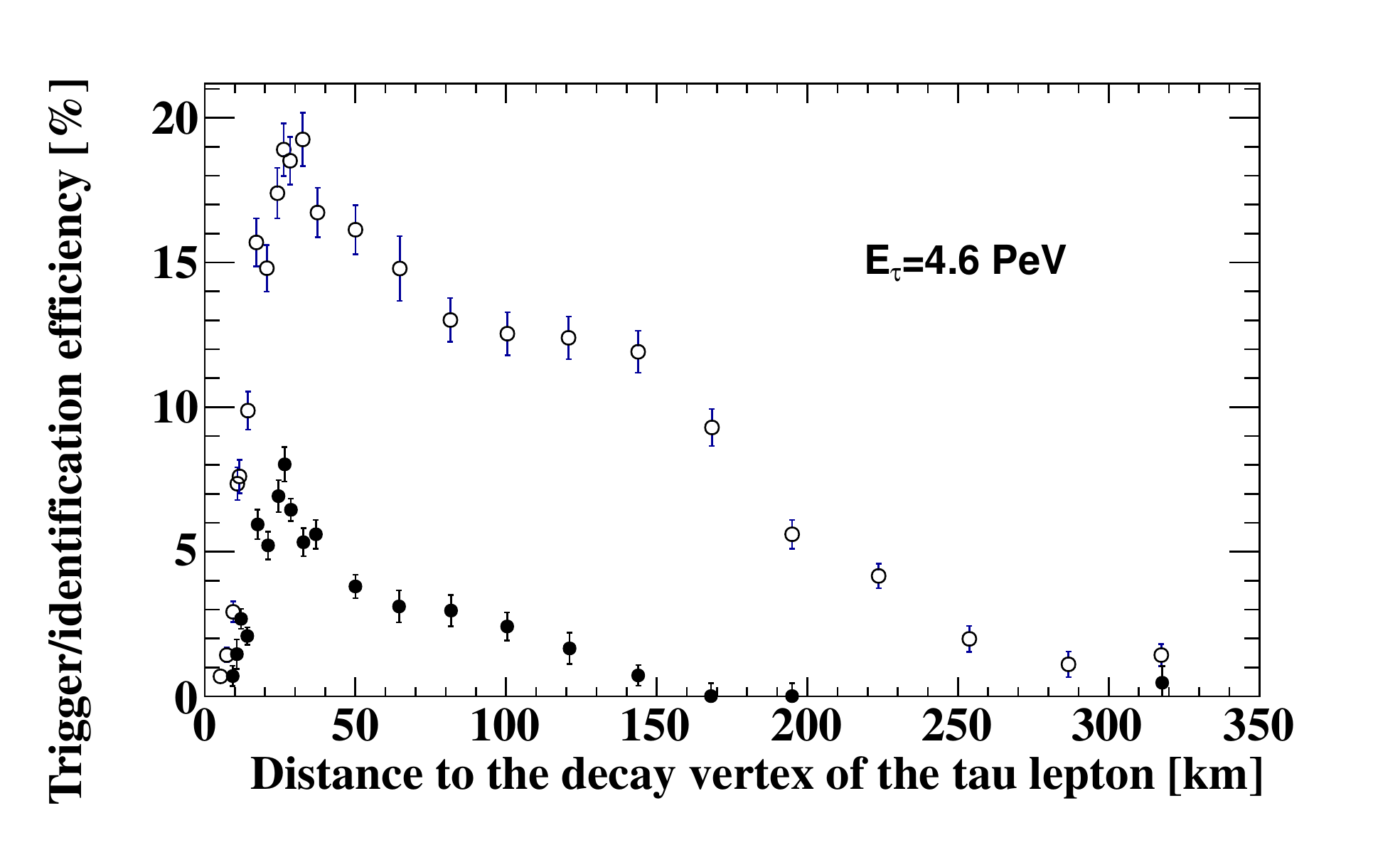}
\includegraphics [width=0.49\textwidth,height=6cm]{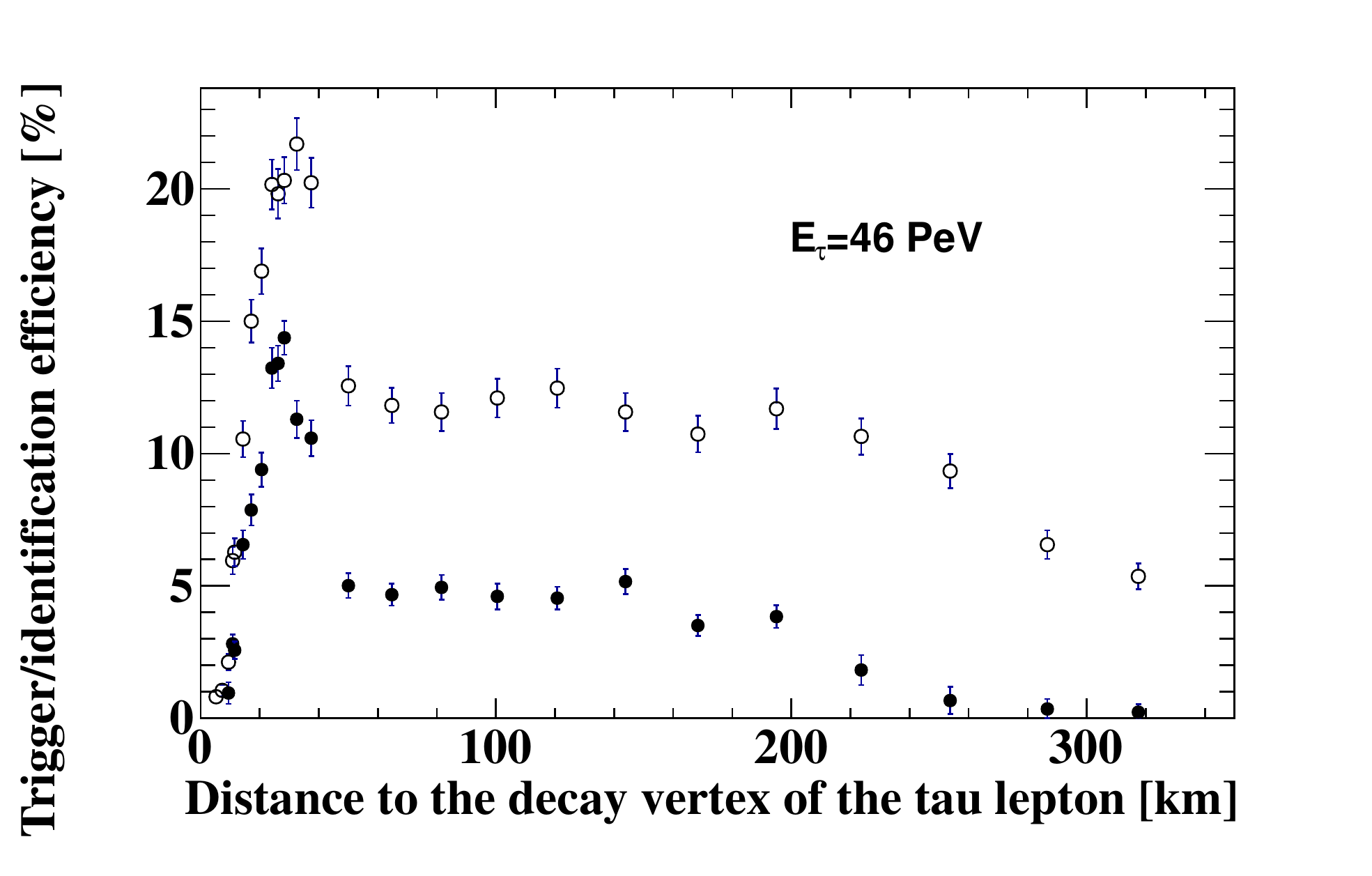}
 \caption{\small Trigger (open circles) and combined trigger and tau lepton identification  efficiency (full circles) for MAGIC as a function of the distance to the decay vertex of the tau lepton, for an  energy of 4.6 PeV (left panel) and 46 PeV (right panel). These efficiencies are calculated  as an average over  simulated   showers  with an impact distance smaller than 1.3 km and the DISTcut applied,  at a zenith angle  $86^{\circ}$. Simulations  are presented for  equal step size in  the vertical depth, i.e. every 20 g/cm$ ^{-2}$.}\label{fig1}
\vspace{-0.25cm}
\end{figure*} 

As already mentioned, high energetic  muons  or   muon bundles can reach the detector and  produce large  shower images.  
The muon bundles are the dominant background contribution of the analysis. A possible contribution from showers induced by cosmic ray electrons can be neglected. At horizontal directions the Cherenkov light from electromagnetic showers will be strongly attenuated, even more than for proton primaries of similar energy, and the shower images in the camera will be too small, see e.g. Figure 4 of \cite{gora:2016}.

In order to calculate the expected number of background events, we used the characterization of the atmospheric muon flux above 
15 TeV measured by IceCube~\cite{icecubemuons}, which can be modeled by an unbroken power law:
\begin{equation} 
\label{eq-bckg}
\frac{d\Phi_{\mu}}{dE_{\mu}} = \phi_{\mu} \times \left( \frac{E_{\mu}}{10 \; \mbox{TeV}} \right)^{-\delta}.
\end{equation}
The values of the parameters that maximize the  likelihood for the parametrized muon flux are: $\phi_{\mu} = 1.06^{+0.42}_{-0.32} \times 10^{-10} \mbox{ TeV$^{-1}$ cm$^{-2}$ sr$^{-1}$s$^{-1}$} $ and a spectral index of $\delta=3.78\pm0.02(\mathrm{stat.})\pm0.03 (\mathrm{syst.}) $. 

This is a conservative assumption, because we do not include  here a dependency on the zenith angle which leads to a lower flux of muons at large zenith angles~\cite{icecubemuons}.   Using the acceptance with height cut shown in Figure \ref{fig2} (right panel) and the  muon flux given by  Eq.  \ref{eq-bckg}, the expected background event rate is at a  level of $4.3 \times 10^{-7}$ events for  one hour of observation.

\subsection{Acceptance of the MAGIC Telescopes}\label{acceptnace:chapter}

The propagation of a given neutrino flux through the Earth and the atmosphere is simulated using  an extended version of the ANIS code~\cite{goraanis}.  The extended version gives a possibility to simulate the  lepton tau propagation in air for different orographic condition of considered site.  For a set of fixed neutrino energies, $10^{6}$ events are generated each on top of the atmosphere for a  zenith angle $\theta=92.5^{\circ}$ and an  azimuth angle $\phi=-30^{\circ}$.
The tau is propagated in small  steps until the age of the tau lepton exceeds the tau lepton lifetime. The different amount of energy loss in the Earth’s crust and air have been also  taken into account, see~\cite{goraanis} for more details.
All computations are done using Digital Elevation Map\footnote{Consortium for Spatial Information (CGIAR-CSI) \url{http://srtm.csi.cgiar.org/}} to model the surrounding mass distribution of the La Palma site. As a results of these simulations, the flux of  leptons emerging from the ground as well as their energy and the decay vertex positions are calculated inside an interaction volume.  The
interaction volume for a given incoming neutrino with energy $E_{\nu}$ is defined by a particular plane $A_{gen}(\theta)$
and distance $\Delta l$,which is a multiple  of a  few times of the average lepton range. The plane $A_{gen}(\theta)$
is also  the cross-sectional area of the detector volume. The detector  volume is modeled by a cylinder  with radius of 50\,km and  10\,km height  with its z-axis (height) pointing upwards, see Figure 3 in~\cite{goraanis}. Since the plane $A_{gen}(\theta)$ was used as reference plane for the generation of incoming neutrinos, by definition, it is orthogonal to the incoming neutrino direction.  

In such an approach the detector  aperture/acceptance 
for an initial neutrino energy $E_{\nu_\tau}$ can be  calculated from:
\begin{eqnarray}
A^{\mathrm{ps}}(E_{\nu_\tau}, \theta,\phi)  =N(E_{\nu_\tau})_{\mathrm{gen}}^{-1} \times \sum_{i=1}^{N_{\mathrm{DISTcut}}} P_{i}(E_{\nu_\tau},E_{\tau},\theta,\phi) \nonumber  \\
    \times  A_{gen,i}(\theta) \times T_{\mathrm{eff},i}(E_{\tau},r_{5 \mathrm{km}},d_{5 \mathrm{km}},\theta),
\label{aperture}
\end{eqnarray}
where $\theta$, $\phi$ are the simulated zenith and azimuth pointing angles of the MAGIC telescope, and  $N_{\mathrm{gen}}(E_{\nu_\tau})$ is the number of generated neutrino events per neutrino energy bin.  The interaction probability is given by $P(E_{\nu_\tau},E_{\tau},\theta,\phi)$,  the probability that a neutrino with energy $E_{\nu_\tau}$ and zenith angle $\theta$  and azimuth angle $\phi$ produces a lepton with energy $E_{\tau}$, which  can reach the detector volume, see again~\cite{goraanis} for more details.

In order to calculate the point source neutrino aperture for the MAGIC, we consider  only   events which 
are in the FOV of the MAGIC. Thus,  $N_{\mathrm{DISTcut}}$ is  the number of $\tau$ leptons with energies $E_{\tau}$ larger than the threshold energy $E_{\mathrm{th}}=1$\, PeV and  after selection cut, which guarantees that  the estimated  position of the shower maximum lies in the FOV of the MAGIC  telescopes. At  its  maximum, a shower has the largest lateral extension and Cherenkov light production, thus is capable of producing the largest signal seen by IACTs telescopes. The following FOV  condition was then used: $\alpha_{5 \mathrm{km}}= \arcsin(d_{5\mathrm{km}}/r_{5 \mathrm{km}})< (\delta_{\mathrm{FOV}}/2+\alpha_{\mathrm{Cher.}}) \simeq 3.10^{\circ}$, where  $d_{5 \mathrm{km}}$ is the  distance  of the estimated shower  maximum to the shower axis, the $r_{5\mathrm{km}}$ is the vector pointing from the telescope to the estimated position of the shower maximum and  $\delta_{\mathrm{FOV}}=3.5^{\circ}$ is  the FOV of the MAGIC camera, (see  Figure~\ref{sketch}). In this selection criterion, called here DISTcut, we ensure that a least good fraction of the shower is imaged into the MAGIC camera, see~\cite{epsvhe2017} for more detailed desription of this cut

 In Eq.¬\ref{aperture}, $T_{\mathrm{eff},i}(E_{\tau},r_{5 \mathrm{km}},d_{5 \mathrm{km}},\theta)$  is the trigger and reconstruction/cut efficiency for $\tau$-lepton induced showers with  its estimated position of the shower maximum  at  distance $r_{5 \mathrm{km}}$ from the telescope  and the distance   $d_{5 \mathrm{km}}$. In case of the  aperture calculations,  Eq. \ref{aperture} was used with $T_{\mathrm{eff},i}$ set to 1, while for the  acceptance  calculations  the trigger and identifications cuts are included i.e. $T_{\mathrm{eff},i} < 1$. 

\begin{figure*}[ht!]
 \centering
 \includegraphics [width=0.49\textwidth,height=7cm]{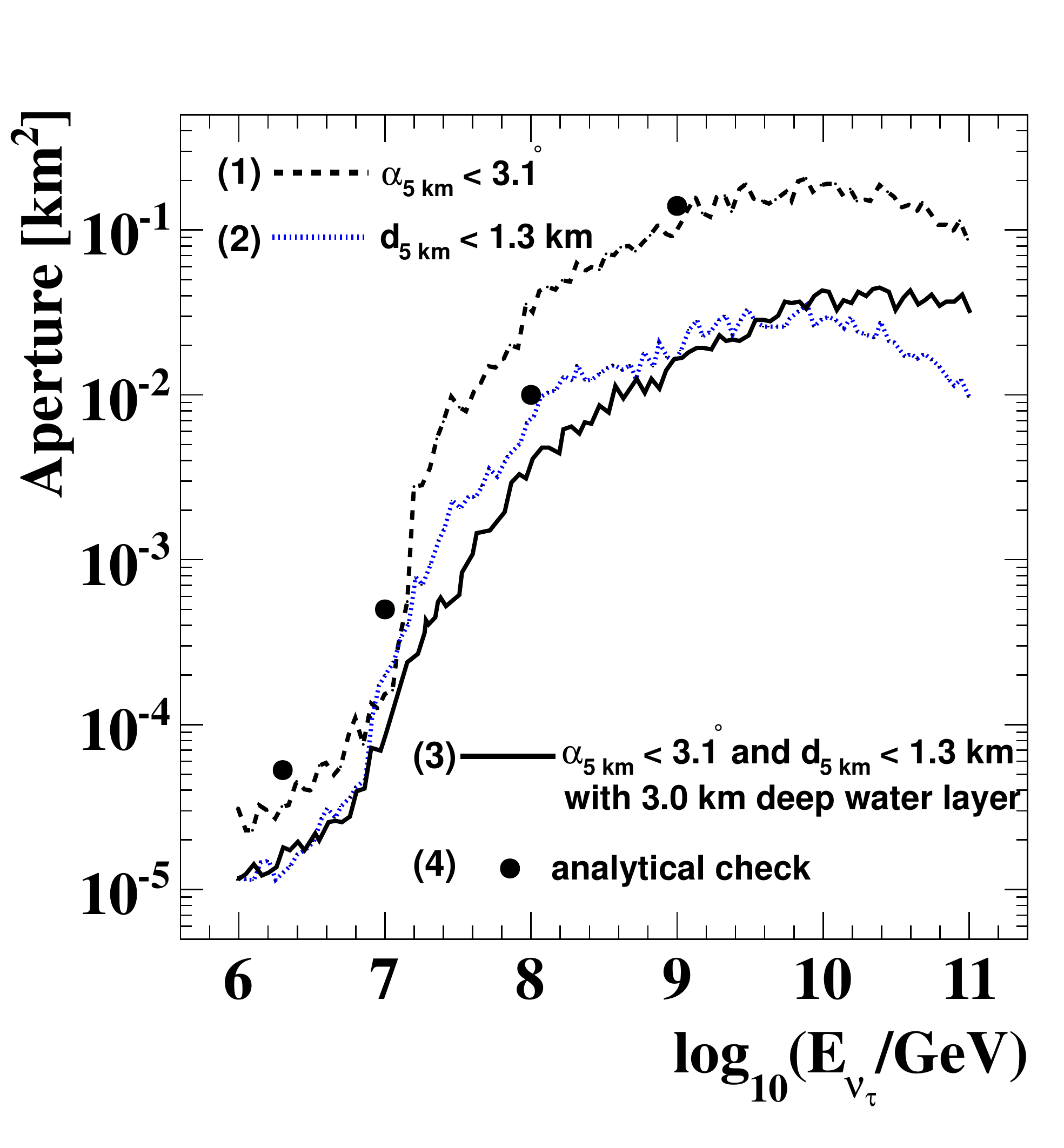}
  \includegraphics [width=0.48\textwidth,height=7cm]{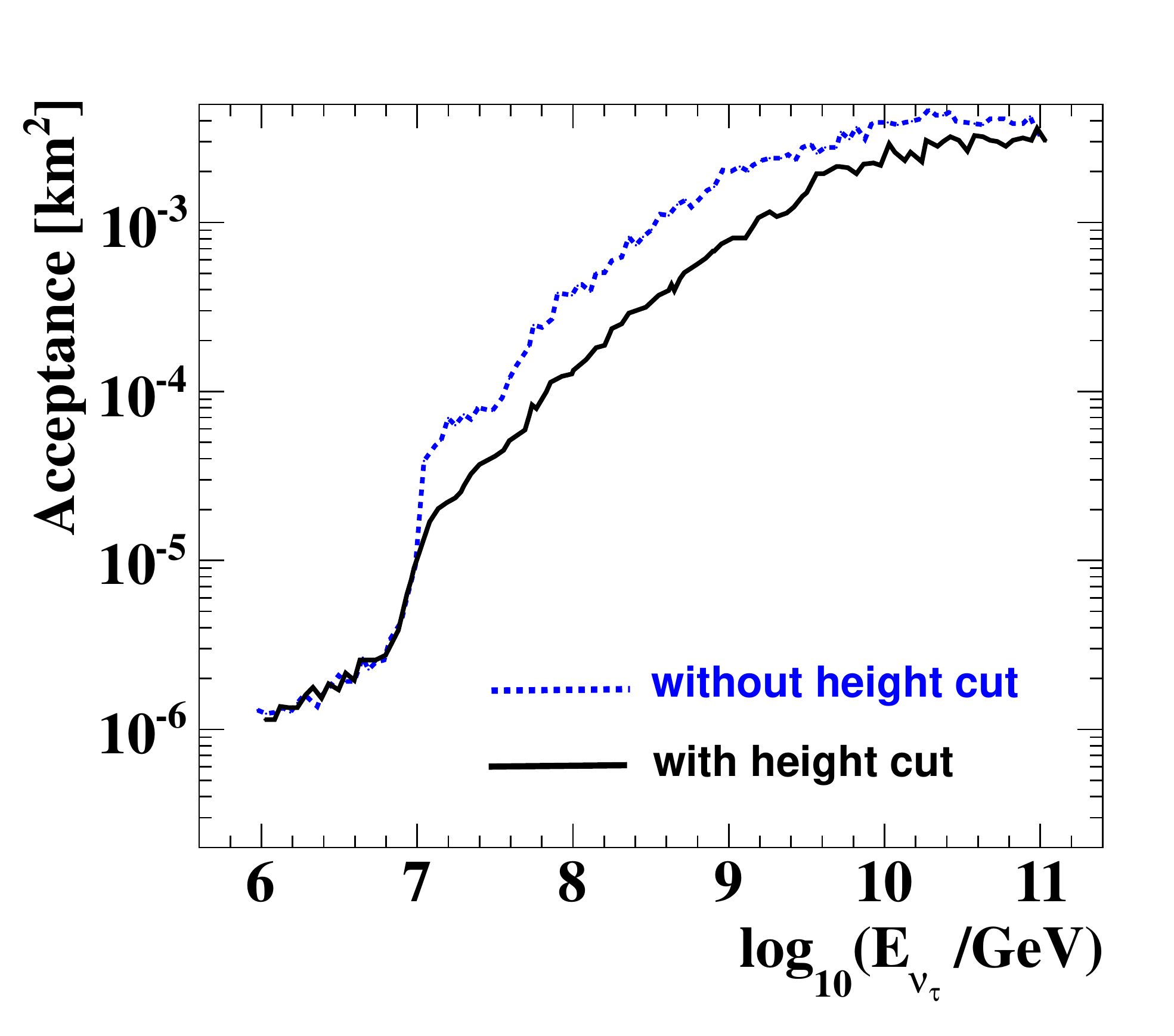}
 \caption{  Left: The point-source aperture   to  earth-skimming tau neutrinos for the MAGIC telescopes pointing at  $\theta=92.5^{\circ}$ and  $\phi=-30^{\circ}$ for different densities of the interaction medium (see text for more details); Right: The acceptance for point sources, $A^{\mathrm{ps}}(E_{\nu_\tau})$  as estimated for the MAGIC site, with a DISTcut of: $\alpha_{5\mathrm{km}}<3.1^{\circ}$ and $d_{5\mathrm{km}} < 1.3$ km, within  identification efficiencies shown in Figure~\ref{fig1} and  with/without the height cut.     }\label{fig2}
\vspace{-0.25cm}
\end{figure*} 

The trigger efficiency depends on the response of a given detector and is usually estimated based on MC  simulations.
 The trigger efficiency in an energy  interval, $\Delta E$, is defined  as the  number of the simulated showers with positive trigger decision over the total number of generated showers for  fixed zenith angle $\theta$, initial energy of the primary particle $E_{\tau}$, and the impact distance. The  impact  distance of simulated showers  was randomized in the  CORSIKA simulations  (by using the CSCAT option)  and later the  Cherenkov telescope orientation  for such showers  was randomized  over the MAGIC camera FOV. In order to evaluate the identification efficiency  for tau neutrino showers we apply in addition the selection  criterion shown in Figure~\ref{fig::dist2} (right panel).
Figure~\ref{fig1} shows an example of the  trigger/identification efficiency for two example energy bins. It is  well seen that for smaller distances ($r_{\mathrm{decay}} < 20$ km) the efficiency drops. In such a case the shower maximum is too close to the detector, and the shower does not reach yet the maximum of shower.
   In general, these plots  provide  an estimate  of the typical distance  for $\tau$-induced showers seen by MAGIC.

In Figure~\ref{fig2} (left panel)   we show  an estimate  of  the MAGIC point-source aperture (for $T_{\mathrm{eff},i}=1$) to tau neutrinos.
The aperture is shown for  four  cases: (1) for  simulations including  the orography of the La Palma island, 
but with the  spherical model of  Earth,  with the rock density of about 2.65 g/cm$^2$, outside the island, and 
 with  $\alpha_{r_{5 \mathrm{km}}}<3.1^{\circ}$; (2)  with an additional  impact distance selection criterion  $d_{5 \mathrm{km}} <1.3$ km; (3)  with  a 3.0~km deep water layer around island~\footnote{A 4 and 4.5 km  deep water layer have been tested as well, which causes a change in the estimated aperture of less than 30\%.}; (4) for aperture calculated using a simple analytical approximation (described in Section~\ref{analytical_check}). 
 The contribution of the water layer is important, leading to a factor two change in the aperture compared to the simulation which includes the orography of La Palma only, case 1).  This is because the $\nu_\tau$ has a much smaller interaction probability in water, and can produce hence smaller escaping $\tau$-lepton fluxes.

An important effect in the analysis is the presence of possible clouds during observations, that needs to be taken into account.
At the MAGIC site due to the location of the detector, two classes of clouds can be found:  one  expected above the MAGIC telescopes 
and other one below. As mentioned in the introduction  the presence of high clouds above the detector can  make impossible normal “gamma-ray” observations but  allow to perform  horizontal observations for tau neutrinos. The MAGIC lidar  system~\cite{lidar} indicates, 
whenever high clouds in the  vertical directions are present at the MAGIC site. This information is usually used to start 
observations at large zenith angles, i.e. observations are performed if the transmission  for the aerosol component  from  9 km to ground   is below 0.55  and  from 3 km  to ground is close to 1.0.   However,   during horizontal observations, we can   also expect  clouds below the MAGIC telescopes. These clouds usually   form  the  quasi-stable layer of cumulus  between 1.5 km and 1.9 km a.s.l. due to the temperature inversion at these altitudes~\cite{Carrillo2016}. For such case,  we did  not have any information about clouds present in the directions of the {\it seaON  } and {\it seaOFF } observations, due to the lack of lidar measurements in these directions\footnote{ With existing  setup, the MAGIC lidar can monitor only clouds layer up to a few tens of kilometers, which is not enough for large zenith angle observations,  where we  need   to know if the clouds are present or not  at much larger distances,  at least  one  hundred of kilometers.}.

Thus, in our acceptance calculations, as  the most conservative case,  we assumed the  presence of the quasi-stable layer of cumulus between 1.5 km and 1.9 km a.s.l., With this assumption all the Cherenkov light generated below 1.9 km is absorbed.
To estimate this effect, all decaying tau leptons below 1.5 km a.s.l. were discarded (referred to as "height cut"). We assume that in this case, the Cherenkov light is absorbed when it crosses the layer between 1.5 and 1.9 km a.s.l. This selection provides a conservative upper limit of this effect. Figure~\ref{fig2} (right panel) shows the acceptance  obtained with and without the "height cut" applied. As we can see in the  plot  this  selection criterion  leads  to a smaller (about factor two) acceptance, showing the influence of the quasi-stable layer of cumulus
and  gives the uncertainty associated to the fact that there are no lidar
measurements in the horizontal direction. After simulating the effects of the orography of the site, the sea and the layer of clouds, and also taking into account the trigger and identification efficiency, we obtain an acceptance which is one or two orders (for $10^{18}$ eV)  of magnitude smaller than calculated aperture.

\begin{table*}[h]
 \caption{\small Comparison of the  aperture ($A^{ps}(E_{\nu_{\tau}})$) from MC simulations (Section~\ref{acceptnace:chapter}) and an analytical approach (Section~\ref{analytical_check}).
  Results are obtained for a zenith angle of $92.5^{\circ}$, neutrino crossing  distance $L=546$ km, an  average rock density $\rho=2.65$ g/cm$^{2}$ and the charged current neutrino cross-section from~\cite{sarkar}. The energy of the tau particle  is approximated as $E_\tau = 0.75 E_{\nu_{\tau}}$.}
 \label{tab:a}
\center
 \begin{tabular}{lccccccc}
 \hline
 \hline
 $E_{\nu_{\tau}}$ & $\sigma_{CC}({E_{\nu_{\tau}}}$) &  $\lambda_{\nu_{\tau}}$ &  $\lambda_{\tau}$  & $P(E_{\nu_\tau},E_{\tau},\theta)$ & $A_{\mathrm{geom}}$   &  $A(E_{\nu_{\tau}})$  & $A^{ps}(E_{\nu_{\tau}})$  \\
  (PeV)&   (pb) & (km) & (km) &    & (km$^2$)     & (km$^2$)  &  (km$^2$) \\
 \hline
\hline
 2  &  950&6596&0.073&$1.0\times 10^{-5 }$ & 5.33 &  $5.3\times10^{-5}$& $1.5\times10^{-4}$\\
 10 & 1900&3298&0.367&$9.4\times10^{-5}$  & 5.33 &  $5.0\times10^{-4}$& $2.7\times10^{-4}$\\
 100& 4800 &1305 &3.670 &$1.9\times 10^{-3}$ & 5.33 & $1.0\times10^{-2}$& $8.0\times10^{-3}$\\
 1000& 11000&569&36.70 &$2.6\times 10^{-2}$ & 5.33 & $1.4\times10^{-1}$& $2.6\times10^{-2}$\\
 \hline
\hline
 \end{tabular}
 \end{table*}



\subsection{Analytical aperture  estimation  of the MAGIC Telescopes}\label{analytical_check}
\begin{figure}[ht!]
 \centering
\includegraphics [width=0.48\textwidth,height=8cm]{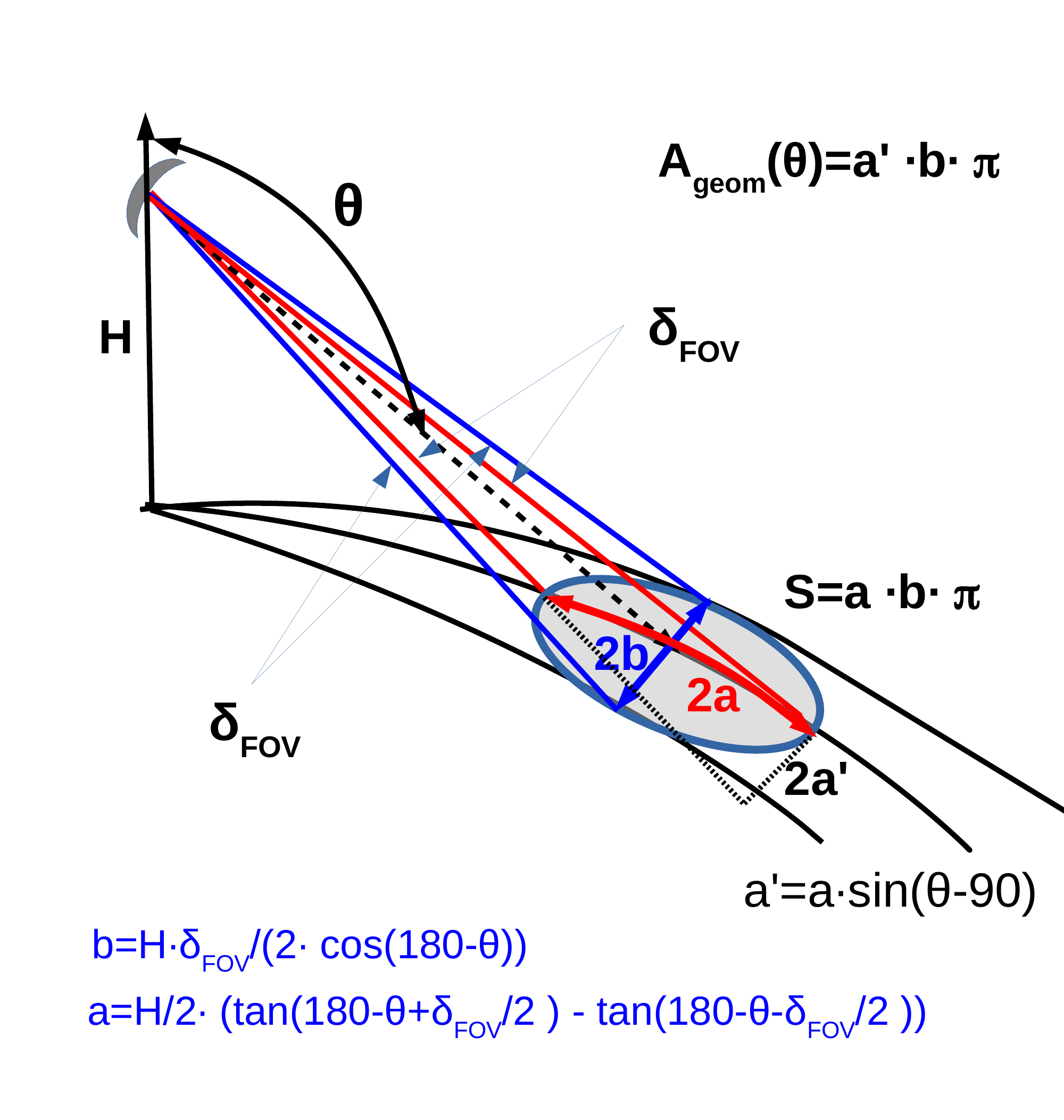}
\caption{\small  The  geometrical area opened by the FOV of the MAGIC telescopes.}\label{elipsa}
\end{figure}

A simple analytical estimate is found to yield the correct order of magnitude for the effective aperture of MAGIC. We focus first on the geometry of the system. As the horizon is observed at a zenith angle of  $\theta > 90^{\circ}$, the particle path  through  the Earth is $L \simeq -2 R_{\mathrm{Earth}} \cdot \cos(\theta)$, where $R_{\mathrm{Earth}}$ is the  Earth radius, see Figure~\ref{fig::sea} (left panel). The geometric area seen by telescopes $A_{\rm geom}$ along the line-of-flight of the neutrino  can be approximated by:
\begin{eqnarray}
A_{\mathrm{geom}}(\theta) = a\cdot b \cdot \pi = H^2 \cdot 
 \frac{\delta_{\mathrm{FOV}}}{4} \cdot \pi \cdot \big(\tan(\theta+\delta_{\mathrm{FOV}}/2) \\ \nonumber
-\tan(\theta-\delta_{\mathrm{FOV}}/2) \big), 
\end{eqnarray}\label{elipsa_area1}
where $a$ and $b$ is the major and minor axis of ellipse  (see Figure~\ref{elipsa}),   $H = 2.2$ km a.s.l. is the altitude  of the telescope and $\delta_{\mathrm{FOV}}=0.061$ rad ($3.5^{\circ}$) is the FOV of the MAGIC camera. The Taylor expansion given in Eq.~4  is accurate within 20\%. 
The ellipse is actually truncated, because the horizon appears already   at
$ \theta_{horizon} = 180^{\circ} - ( 180^{\circ} / \pi \times \arcsin(R_\mathrm{Earth}/(R_\mathrm{Earth}+H)) ) = 91.5$ deg. This  effect reduces the geometrical area  $A_{\mathrm{geom}}(92.5^{\circ})$ from 14.3 km$^2$  to about 5.33 km$^2$.

The conversion efficiency  for tau neutrinos  along the distance $L$ is  calculated from~\cite{huang}:
\begin{eqnarray}
P(E_{\nu_\tau},E_{\tau},\theta)= 
\int_0^L \exp{\left(-x/\lambda_{\nu_{\tau}}\right)}\cdot \exp{\big(-(L-x)/\lambda_{\tau}\big)} \cdot
		\frac{dx}{\lambda_{\nu_{\tau}}} \\ \nonumber
            =  \frac{\lambda_{\tau}}{\lambda_{\nu_{\tau}} - \lambda_{\tau}} \cdot
	\big( \exp{\left(-L/ \lambda_{\nu_{\tau}}\right)} - \exp{\left(-L/ \lambda_{\tau}\right)} \big), 
\label{eq:convert}
\end{eqnarray}
where $\lambda_{\nu}=1/(\sigma_{\textrm{CC}}N_{A}\rho)$ is the neutrino mean free path,
 $N_{A}$ the Avogadro constant, $\sigma_{\textrm{CC}}$ the charged current neutrino interaction  cross-section ~\cite{sarkar} and $\rho$ the density of the rock.
The  decay length of the  tau particle  $\lambda_{\tau}=48910 \mbox{ } \mathrm{m} \times (E_{\tau}/{ 1000 \mbox{ } \mathrm{ PeV}})$. 
The effective area  can be estimated as  $A(E_{\nu_\tau},\theta) =  P(E_{\nu_\tau},E_{\tau},\theta) \times A_{\mathrm{geom}}(\theta)$,
under the approximation that all tau leptons decay close to  the sea surface, which is  true  only for tau lepton energies below $\sim$30 PeV, and without taking into account trigger and cut efficiencies.
 If a tau lepton above an energy of a few tens of PeV still escapes  the sea, its decay length can be  too large to initiate an air shower before reaching
the telescopes.
 In Table~\ref{tab:a}, the conversion efficiencies and comparison of the effective  area ($A(E_{\nu_\tau})$) with our MC estimate ($A^{ps}(E_{\nu_\tau})$) are shown for five energies. The simple analytical calculations agree
 (apart from the last point at 1000 PeV),  with the aperture estimate from MC simulations  within a factor of 2  (see  Figure ~\ref{fig2}, left).  
  
\begin{table*}[h]
\caption{\small  Expected event rates in the MAGIC telescopes for AGN flares with the Flux predictions described in Section \ref{agnflux}.}\label{table3}
\center
\begin{tabular}{cccccccccc}
\hline
\hline
& &\bf Flux-1  &\bf Flux-2& \bf  Flux-3 & \bf Flux-4 & \bf Flux-5 \\
& & \bf ($\times 10^{-5}/3$ hrs)   & \bf ($\times 10^{-5}/3$ hrs)  &  \bf ($\times 10^{-5}/3$ hrs)    &\bf ($\times 10^{-5}/3$ hrs)  &\bf ($\times 10^{-5}/3$ hrs)   \\
\hline
\hline
$N_{\mathrm{Events}}$& without height cut  & 2.4 & 1.4 & 0.74   &7.4 &2.4\\
$N_{\mathrm{Events}}$& with height cut  & 1.1&  0.6 &  0.30  & 2.9 & 1.2 \\
\hline
\hline
\vspace{-0.7cm}
\end{tabular}
\end{table*} 

	. 
\subsection{Event rates}
 The total number of expected signal events N is obtained as:
\begin{equation}
N=\Delta T \times \int_{E_{\mathrm{th}}}^{E_{\mathrm{max}}} A^{\mathrm{ps}}
(E_{\nu_\tau})\times\Phi(E_{\nu_\tau})\times dE_{\nu_\tau},
\end{equation}
 where  $\Delta T$  is the observation time, $A^{\mathrm{ps}}(E_{\nu_\tau})$ the  point source acceptance and $\Phi(E_{\nu_\tau})$ the expected neutrino flux.  Since  $A^{\mathrm{ps}}(E_{\nu_\tau})$ depends on the zenith angle and hence on time (since all sources move in the sky),  in reality, an integral from 0 to $\Delta T$ must be made over time $t$, and  $A^{\mathrm{ps}}(E_{\nu_\tau})$  should be replaced  by $ A^{\mathrm{ps}}(E_{\nu_\tau}, t)$. Thus,  all numbers presented in the following have to be considered  approximations since they used a time-independent value of   $A^{\mathrm{ps}}(E_{\nu_\tau})$.

  Here  we provide an estimate of the event rate for a sample  of generic neutrino fluxes, from photo-hadronic interactions in case of flaring AGNs, if observed at the most efficient zenith angle. Flaring  AGNs can produce a boosted flux of neutrinos.
 Table~\ref{table3} shows  the expected event rates for MAGIC, using the flux  benchmark models  shown  in Figure~\ref{fig::spectrum2}. 

 The rates are calculated for tau neutrinos   assuming that the source is in  the MAGIC  FOV  for a period of 3 hours and with the acceptance calculated with and without the ''height cut'' applied. In the case of Flux-3 and Flux-4 for events with energies of the $\sim$$10^{8}$ GeV, the expected event rate with the "height cut" applied is of the order $3 \times 10^{-5}$ events per 3 hours. In the case of Flux-1, Flux-2 and Flux-5, for events below $\sim$$ 5 \times 10^{7}$ GeV, the number of expected events is below $1.1 \times 10^{-5} $ in 3 hours. In case the "height cut" selection is not applied, the expected number of events increases by a factor 2. We give results for both cases, because  sometimes it is also
possible that observations can be performed during absence of quasi-stable layer of cumulus in {\it seaON} direction.

\begin{table}[h]
 \caption{\label{tab::rate} {Relative contributions to the systematic uncertainties on the up-going tau neutrino rate. 
 Systematic uncertainty on the expected tau neutrino rate due the neutrino-nucleon cross section and the tau-lepton energy loss. Both uncertainties have been added in quadrature. As a reference GRV98lo and ALLM model for Flux-1 and Flux-3 was used. }}
\center 
\begin{tabular}{ccccc}
\hline
\hline
model &  cross-section  & $\beta_{\tau}$ & Total    \\
\hline
\hline
Flux-1 
& $^{+14\%}_{-2\%}$    
& $^{+2\%}_{-7\%}$   
& $^{+14\%}_{-7\%}$ \\
& & &\\
Flux-3 
&  $^{+42\%}_{-7\%}$   
&  $^{+7\%}_{-14\%}$ 
& $^{+43\%}_{-16\%}$\\
\hline
\hline
\end{tabular}
\end{table}

The systematic uncertainties on the event rates due to the tau-lepton energy loss and the neutrino-nucleon cross section have been taken into account.  The average energy loss of tau particles per distance travelled (unit depth $X$ in g\,cm$^{-2}$) can be described as $\left\langle dE/dX \right \rangle = \alpha(E) + \beta(E) E$. The factor $\alpha(E)$, which is nearly constant, is due to ionization. $\beta(E)$ is the sum of $e^+e^-$-pair production and bremsstrahlung  and photonuclear scattering, which is not only the dominant contribution at high energies but at the same time subject to relatively large uncertainties. In this work, the factor $\beta_{\tau}$ is calculated using the following models describing the  contribution of photonuclear scattering: ALLM~\cite{allm}, BB/BS~\cite{bbbs}, CMKT~\cite{ckmt}, and different neutrino-nucleon cross-sections: GRV98lo~\cite{GRVlo}, CTEQ66c~\cite{cteq}, HP~\cite{hp}, ASSS~\cite{sarkar}, ASW~\cite{Albacete:2005ef}. The  results are listed in~Table~\ref{tab::rate} for Flux-1 and Flux-3, and show that the combined systematic uncertainty of both effects can be considerable, namely of the order of 40\%, but nevertheless much smaller than the effect of the quasi-stable layer of cumulus below the observatory, included in the ''height cut'', which can lead to a factor two of the lower event rate. 

 \begin{figure*}[h]
 \centering
 \includegraphics [width=0.55\textwidth]{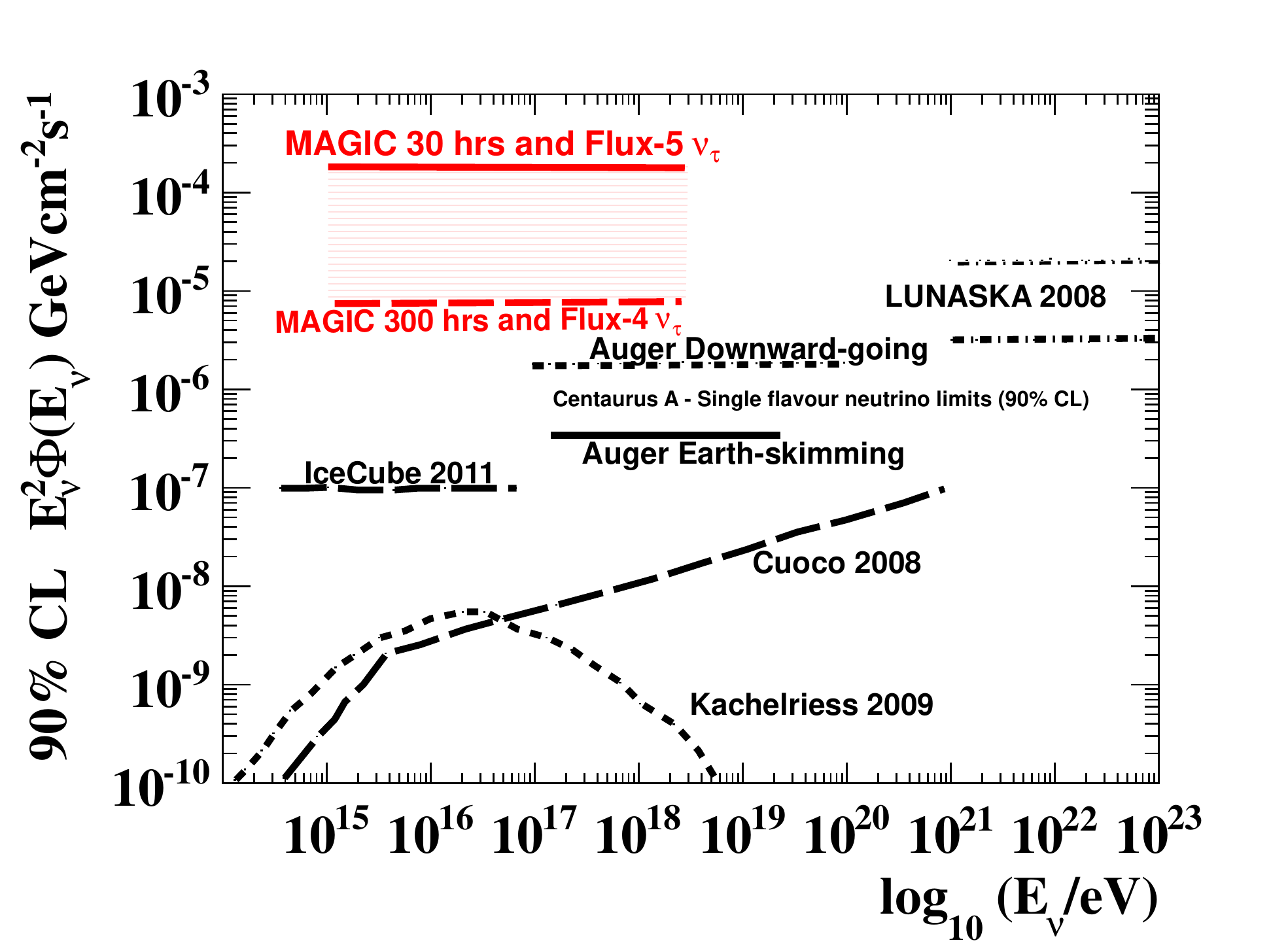}
 \caption{\small 90\% C.L. upper limit on the tau neutrino flux obtained with the MAGIC telescopes with 30 hrs of observation (red solid line) assuming Flux-5 (see Figure \ref{fig::spectrum2}). The expected upper limit with 300 hrs of observations assuming Flux-4, is shown as the red dashed line. The results are compared to the 90\% C.L. upper limit on the single flavour neutrino flux from Pierre Auger~\cite{auger},  IceCube  ~\cite{icecube2011} and LUNASKA 2008~\cite {lunaska}. The predicted fluxes for two theoretical models of ultra high neutrinos   production – in the jets ~\cite{cucco} and close to the core of Centaurus A~\cite {kache}– are also shown for comparison. Plots adopted from ~\cite{auger}.}\label{fig111}
\end{figure*} 

\section{Tau neutrino flux limit}
From the estimated acceptance with height cut, the sensitivity for an injected spectrum $K\times\Phi(E_{\nu})$ with a known shape $\Phi(E_{\nu})$ was  calculated. As no events survived after event selection, 90\% C.L. upper limits 
~\cite{limit}  on the tau neutrino flux have been obtained. Assuming a reference spectrum of  $\Phi(E_{\nu})=1 \times 10^{-8} E^{-2} \mbox{ GeV$^{-1}$ cm$^{-2}$ s$^{-1}$}$ of a point-like source, the upper limit obtained is:
 $K_{90\%}=2.44/N_\mathrm{Events}$.  The limit for a point source search  is then:
\begin{equation}
 E_{\nu_\tau}^{2}\Phi^{ps}(E_{\nu_\tau}) < 2.0   \times 10^{-4} \mbox{ GeV cm$^{-2}$ s$^{-1}$} \label{limit}
\end{equation}
 where $E_{\nu_\tau}$ is in the range between 1  and 3000 PeV. 
 The neutrino flux upper limit is obtained for an expected number of tau neutrino events of $N_{\rm Events}=1.2 \times 10^{-4}$, in the case of Flux-5, and is shown in Figure 11 (solid red line). The result is also compared to the 90\% C.L. upper limit on the single flavor neutrino flux from the Pierre Auger experiment~\cite{auger} from the active galaxy Centaurus A. The expected MAGIC limit could be improved in the case of 300 hours of observations during a strong flare as in Flux-4, where  a limit of $E_{\nu_\tau}^{2}\Phi^{ps}(E_{\nu_\tau}) < 8.4\times 10^{-6}$ \mbox{ GeV cm$^{-2}$ s$^{-1}$} can be obtained. This expectation is shown in Figure~\ref{fig111} as the dashed red line, and is only a factor 3.4 worse than the Pierre-Auger "down-going" analysis.

\section{Summary}

In this paper, a search for tau neutrinos of astrophysical origin in the energy range between 1 PeV and  3 EeV with the MAGIC telescopes is presented. The data was collected during a special pointing of the telescopes below the horizon, to detect Earth-skimming tau-lepton induced showers. These observations can take place during periods of high clouds, which prevent standard gamma ray observations. A 90\% C.L. upper limit on the tau-neutrino flux of
$ E_{\nu_\tau}^{2}\Phi^{ps}(E_{\nu_\tau}) < 2.0   \times 10^{-4}$  GeV cm$^{-2}$ s$^{-1}$ was obtained, with 30 hours of observation. 
The limit  is not competitive with other experiments, hovewer  to our knowledge this is  first time that is has been calculated  with realistics assumptions and using backround data collected by MAGIC.  Thus  our search  gives a realistic illustration of the potential  of the Cherenkov technique for this  present active topic of research. The presented results   can also be important for future Cherenkov experiments like for example  the Cherenkov Telescope Array.
This next generation ground-base observatory can have a much better possibility to detect tau neutrinos, given its a larger FOV (e.g. in extended observation mode) and much larger effective area

\section{Acknowledgments}

%
%
We would like to thank the Instituto de Astrof\'{\i}sica de Canarias for the excellent working conditions at the Observatorio del Roque de los Muchachos in La Palma. The financial support of the German BMBF and MPG, the Italian INFN and INAF, the Swiss National Fund SNF, the ERDF under the Spanish MINECO (FPA2015-69818-P, FPA2012-36668, FPA2015-68378-P, FPA2015-69210-C6-2-R, FPA2015-69210-C6-4-R, FPA2015-69210-C6-6-R, AYA2015-71042-P, AYA2016-76012-C3-1-P, ESP2015-71662-C2-2-P, CSD2009-00064), and the Japanese JSPS and MEXT is gratefully acknowledged. This work was also supported by the Spanish Centro de Excelencia ``Severo Ochoa'' SEV-2012-0234 and SEV-2015-0548, and Unidad de Excelencia ``Mar\'{\i}a de Maeztu'' MDM-2014-0369, by the Croatian Science Foundation (HrZZ) Project IP-2016-06-9782 and the University of Rijeka Project 13.12.1.3.02, by the DFG Collaborative Research Centers SFB823/C4 and SFB876/C3, the Polish National Research Centre grant UMO-2016/22/M/ST9/00382 and by the Brazilian MCTIC, CNPq and FAPERJ.

\end{document}